\documentclass[11pt,twoside]{article}

\usepackage{amsmath,amsfonts,amssymb,amsthm}
\usepackage{latexsym}
\usepackage{epsfig,overpic}
\usepackage{stmaryrd}
\usepackage{color}
\usepackage{graphicx}
\usepackage[margin=1in]{geometry}
\usepackage{cite}

\newcommand{\bu}{\mathbf{u}}
\newcommand{\bI}{\mathbf{I}}
\newcommand{\ep}{\varepsilon}
\newcommand{\bv}{\mathbf{v}}
\newcommand{\bw}{\mathbf{w}}

\newcommand{\bx}{\mathbf{x}}

\newcommand{\bn}{\mathbf{n}}

\newcommand{\bs}{\mathbf{s}}
\newcommand{\dx}{{\rm d}\bx}
\newcommand{\ds}{{\rm d}\bs}
\newcommand{\blf}{\mathbf{f}}
\newcommand{\Div}{\operatorname{\bf div}}

\date{}

\begin{document}

\title{Finite stopping times for freely oscillating drop of a yield stress fluid\thanks{Partially supported by NSF through the Division of Mathematical Sciences grants 1315993 and 1522252.}}
\author{Wanli Cheng\thanks{Department of Mathematics, University of Houston, Houston, Texas 77204-3008; {\tt wanli@math.uh.edu}} \and
        Maxim A. Olshanskii \thanks{Department of Mathematics, University of Houston, Houston, Texas 77204-3008;
{\tt molshan@math.uh.edu}}
}

\newtheorem{Theorem}{\quad Theorem}[section]

\newtheorem{Definition}[Theorem]{\quad Definition}

\newtheorem{Corollary}[Theorem]{\quad Corollary}

\newtheorem{Lemma}[Theorem]{\quad Lemma}
\newtheorem{remark}[Theorem]{Remark}
\newtheorem{Example}[Theorem]{\quad Example}
\newtheorem{Proposition}[Theorem]{\quad Proposition}

\maketitle

\begin{abstract}
The paper addresses the question if there exists a finite stopping time for an unforced motion of a yield stress fluid with free surface. A {variational} inequality formulation is deduced  for the problem of yield stress fluid dynamics with a free surface. {The} free surface is assumed to evolve with a normal velocity {of} the flow. We also consider capillary forces acting along the free surface. Based on the variational inequality formulation an energy equality is obtained, where kinetic and free energy rate of change is in a balance with the internal energy viscoplastic dissipation and the work of external forces. Further, the paper considers free small-amplitude oscillations of a droplet of Herschel-Bulkley fluid under the action of surface tension forces. Under certain assumptions it is shown that the finite stopping time $T_f$ of oscillations exists once the yield stress parameter is positive and the flow index $\alpha$ satisfies $\alpha\ge1$. Results of several numerical experiments illustrate the analysis, reveal the dependence of $T_f$ on problem parameters and suggest an instantaneous transition of the whole drop from yielding state to the rigid one.
\end{abstract}


\section{Introduction}

There are many materials in nature and industry exhibiting the viscoplastic behaviour.
For example, these are fresh concrete, geomaterials, colloid solutions, powder mixtures,
lubricants, metals under pressure treatment, blood in a capillary, foodstuffs, toothpaste.
Such {a} medium below a certain stress value behaves as a rigid body and above
this level behaves as an incompressible fluid. In many applications such as
geophysical hazards  (e.g., \cite{ancey2007plasticity,griffiths2000dynamics}) or the damping of water waves by a muddy bottom
\cite{tsamopoulos2008steady},  the complex dynamics of viscoplastic  fluids is coupled to the evolution a free surface.
Recently there has been a significant increase of interest
in developing and analyzing mathematical models and numerical methods for  flows of yield stress fluids,
flows with free surfaces and a combination thereof. However, ``yield stress fluids slowly yield to analysis'' \cite{bonn2009yield}.
This paper contributes to finding an answer to the following question:
If there exists a finite stopping time $T_f$ for a free-surface flow of an isolated volume of a yield stress fluid with surface tension forces?
 If the answer is positive, one is also interested in knowing an estimate of $T_f$.

The property of an unforced yield stress fluid flow to come to a complete rest in a finite time is an intrinsic one and often considered of keen importance\cite{balmforth2014yielding}.
For the yield-stress  fluid  flows in pipes of a constant cross-section this property of a weak solution to the variational formulation of the problem has been proved in \cite{glowinski2008lectures,huilgol2002variational}. Theoretical upper bounds for the finite stopping times of several simple one-dimensional flows can be found in \cite{huilgol2015fluid,huilgol2002finite,muravleva2010unsteady}.
In the presence of a free surface, one may distinguish between the existence of a finite cessation time and the existence of a final arrested state (the latter can be attained in a finite or infinite time). Although,
there is a common {belief} that the yield stress  should bring an unforced free-surface flow to rest in a finite time, we are not aware of { a mathematical analysis of this phenomenon except a few special flows}.  The question is intriguing since the theory for viscoplastic \textit{films} with a free surface suggests infinite stopping times \cite{matson2007two}. This may be an artifact of the thin-film approximation. 

The problem of yield stress fluid dynamics with free surface has been addressed also numerically.
Since the full problem poses a serious challenge for numerical simulations,  it is common in the literature to consider simplified models of free-surface yield stress fluids. The shallow approximation is one of the most common reduced model for viscoplastic fluids flows  over {inclined} planes and more complex 2D topographies, see  \cite{ancey2009dam,balmforth2006viscoplastic,hogg2009slumps} for recent reviews on this subject and \cite{acary2012well,bernabeu2014numerical,fernandez2014efficient,ionescu2013viscoplastic} for more recent  advances. The previous studies of free surface viscoplastic fluid flows also include  axisymmetric squeezing flows,  bubble  Bingham type flows~\cite{karapetsas2006transient,alexandrou2013squeeze,tsamopoulos2008steady},  the free interface lattice Boltzmann model~\cite{ginzburg2002free}, and the dam-break problem~\cite{nikitin2011numerical}.

The present paper first considers a full 3D model of vicoplastic fluid flow with a free-surface and surface tension forces.
We deduce a suitable variational inequality formulation  satisfied by any sufficiently regular solution to the fluid model.
As usual, the variational inequality provides us with an energy balance. However, contrary to the wall-bounded flows,
the free energy (due to surface tension) enters the energy balance and makes the analysis harder. This {and the lack of the embedding of $W^{1,1}(\Omega)$ in $L^2(\Omega)$ for $\Omega\in\mathbb{R}^3$ do not permit} us to give the affirmative answer to the question raised above  { for a general free-surface viscoplastic fluid flow (cf. Remark~\ref{rem1})}.
To gain more insight into the phenomenon, we further consider the problem of motion for a viscoplastic drop for which the evolution is driven only by surface tension forces. { Droplet flows of yield stress fluids, such as molten metals or polymers, arise in many engineering applications, including spray coating, 3D printing and arc welding~\cite{aziz2000impact,haidar1996predictions,coussot2014yield}. In these and some other applications, surface tension forces play essential role in the formation and evolution of fluid droplets, see, e.g., \cite{coussot2014yield,saidi2010influence}. Thus, the oscillating viscoplastic droplet problem is also of its own interest as a model problem for such industrial flows.}

Following the classical analysis of the Newtonian case {\cite{Lamb,reid1960oscillations,miller1968oscillations,prosperetti1980free}}, we assume that the initial shape of the drop is a \textit{perturbation} of a sphere.
For the \textit{Newtonian} fluid, a linear stability analysis predicts that the drop oscillates, while an amplitude of the oscillations decays exponentially to zero
with a damping factor depending on the viscosity.  To the best of our knowledge, this problem has never been analyzed for a viscoplastic fluid.
Under certain assumptions the analysis in this paper shows that in the presence of the {yield stress} the oscillations cease in a finite time $T_f$.  
In a series of numerical experiments we study the dependence of $T_f$ on problem parameters: yield stress, flow index, and  viscosity  coefficient.

The rest of the paper is organized as follows. Section~\ref{s_model} recalls the mathematical model. The variational inequality
formulation is deduced in section~\ref{s_var}.  After some necessary preliminaries section~\ref{s_energy} derives the energy balance for the problem. In section~\ref{s_drop}, we study free oscillations of a viscoplastic fluid droplet. Here we restrict the analysis to the case of the Herschel-Bulkley fluid with {a} fluid index $\alpha \ge 1$, which includes the classic Bingham
 fluid for $\alpha=1$. Results of several numerical experiments in section~\ref{s_num} illustrate the analysis and reveal
 some further interesting properties of the problem. In particular, it shows an instantaneous transition of the whole drop from fluidic state to the rigid one, when the motion ceases.

\section{Mathematical Model} \label{s_model}
Consider the motion of an incompressible fluid  with free surface. Assume that the fluid occupies a bounded time dependent domain $\Omega(t) \in \mathbb{R}^3$ for $t\in[0,\infty)$.
Denote the boundary of domain $\Omega(t)$ by $\Gamma(t)$. One can distinguish between the static boundary (walls) $\Gamma_D$
and a free surface part $\Gamma_f(t)$ so that  $\overline{\Gamma(t)}=\overline{\Gamma_D} \cup \overline{\Gamma_f(t)}$.
In this paper, the whole boundary is assumed to be the free surface, i.e. ${\Gamma(t)}= {\Gamma_f(t)}$ for all $t\in[0,\infty)$.
The analysis of sections~\ref{s_var} and~\ref{s_energy} can be easily extended to the case of $\Gamma_D\neq0$.
The conservation of mass and momentum is given by the system of equations
\begin{equation}\label{N-S}
\left\{
\begin{split}
\rho (\frac{\partial \mathbf{u}}{\partial t}+(\mathbf{u}\cdot  \nabla \mathbf{u}))- \Div \boldsymbol{\sigma} =\blf \\
\nabla \cdot \mathbf{u} =0
\end{split}\right.
\quad\text{in}~~\Omega(t),
\end{equation}
where $\mathbf{u}$ is the fluid velocity field,   $\boldsymbol{\sigma}$ is the stress tensor,
$\blf$ are given external forces, $\rho$ is the fluid density coefficient.

The most important feature of a viscoplastic fluid is its yield stress: {Once the stresses exceed a positive threshold
parameter, the material flows like a fluid. Otherwise,} it behaves like a solid.
To account for such a two-fold behaviour, one imposes conditioned constitutive relations between the strain-rate tensor
$\mathbf{Du} = \frac{1}{2}[\nabla\bu + (\nabla\bu)^T]$ and $\boldsymbol{\tau}$, the deviatoric part of the stress tensor, $\boldsymbol{\sigma}_{ij}=\boldsymbol{\tau}_{ij}{-}p\delta_{ij}$, with pressure $p$. One common choice is  the Herschel-Bulkley constitutive law:
\begin{equation}\label{eq_const}
\begin{aligned}
 \boldsymbol{\tau}=(K|\mathbf{\mathbf{Du}}|^{\alpha-1}+\tau_s |\mathbf{\mathbf{Du}}|^{-1})\mathbf{\mathbf{Du}}~~
  \Leftrightarrow~~  |\boldsymbol{\tau}| >\tau_s& \\
\mathbf{\mathbf{\mathbf{Du}}}=0~~ \Leftrightarrow~~   |\boldsymbol{\tau}|  \le \tau_s&
\end{aligned}
\end{equation}
where   $\tau_s$ is the yield stress parameter, $K$ is the consistency parameter,
$\alpha>0$ is the flow index (for $\alpha<1$ the fluid exhibits shear-thinning property, whereas for
 $\alpha>1$ it is shear-thickening;   $\alpha=1$ corresponds to the classic case of the Bingham plastic).
Further notations introduced above are the following: For a tensor $\mathbf{A}$, $|\mathbf{A}|$ denotes its {Frobenius} norm
\[ |\mathbf{A}|=(\mathbf{A}:\mathbf{A})^{\frac{1}{2}}=(\sum_{1\le i,j \le 3}|A_{ij}|^2)^{\frac{1}{2}},\] $\Div$ denotes
 the vector divergence operator.
Thus, the constitutive relations \eqref{eq_const} imply that the medium deforms and
shows fluidic behavior for $|\boldsymbol{\tau}| >\tau_s$; while for  stresses not exceeding $\tau_s$ the medium either moves as a rigid body or is at rest.


At the initial time $t=0$ the domain and the velocity field are known,
\begin{equation}
\Omega(0)=\Omega_0,\quad  \bu|_{t=0}=\bu_0,\quad \nabla \cdot \bu_0 =0.
\end{equation}
For $t>0$ we assume that the free surface $\Gamma (t)$ is passively advected by the fluid, i.e. the following
kinematic  condition  is satisfied:
\begin{equation}\label{eq_vu}
v_{\Gamma}=\mathbf {u \cdot n} \quad \text{on}~~ \Gamma(t),
\end{equation}
where $\mathbf{n}$ is the {unit external} normal vector on $\Gamma(t)$ and $v_{\Gamma}$ is the normal velocity of $\Gamma(t)$.
Another boundary condition on $\Gamma(t)$ results from balancing the surface tension forces and the fluid stress forces:
\begin{equation}\label{b2}
\boldsymbol{\sigma} \bn|_{\Gamma}= -\gamma \kappa \bn - p_{ext} \bn  \quad  \text{on}~~ \Gamma(t),
\end{equation}
where $\kappa$ is the sum of principal curvatures, $\gamma$ is the surface tension coefficient, $p_{ext}$ is
an exterior pressure which we set to be zero for the rest of the paper, $p_{ext}=0$.

The system of equations, boundary and initial conditions \eqref{N-S}--\eqref{b2} forms a  mathematical
formulation of the problem of  the  Herschel-Bulkley incompressible fluid flow with free-surface. The
problem is challenging for analysis and only partial results are known regarding
its well-posedness, see, e.g. \cite{bulicek2012unsteady,eberlein2012existence} and the reference therein for analysis of wall-bounded Herschel-Bulkley flows. In the next section we show that any solution to \eqref{N-S}--\eqref{b2} (if it possesses certain smoothness) satisfies a variational inequality.

\section{Variational inequality} \label{s_var}

For arbitrary smooth divergence-free vector field $\mathbf{v}$, we first take the scalar product of the
first equation in \eqref{N-S} with $\mathbf{\mathbf{v-u}}$. This gives the relation
\begin{equation}\label{inner product}
\rho (\frac{\partial \mathbf{u}}{\partial t}+(\mathbf{u}\cdot\nabla \mathbf{u}))\cdot(\mathbf{\mathbf{v-u}})-
\Div  \boldsymbol{\sigma} \cdot (\mathbf{\mathbf{v-u}})
=\blf\cdot(\mathbf{\mathbf{v-u}})\quad\text{on}~\Omega(t),
\end{equation}
for all $t>0$.
Further we integrate \eqref{inner product} over  $\Omega(t)$ and obtain after integration by parts the identity
\begin{equation*}
\int_{\Omega(t)}\left\{\rho(\frac{\partial \mathbf{u}}{\partial t}+(\mathbf{u}\cdot\nabla \mathbf{u}))(\mathbf{\mathbf{v-u}})
+ \boldsymbol{\sigma} : \nabla(\mathbf{\mathbf{v-u}}) -\blf\cdot(\mathbf{\mathbf{v-u}})\right\}\dx=
\int_{\Gamma(t)}\boldsymbol{\sigma} (\mathbf{\mathbf{v-u}})\cdot \mathbf{n}\,\ds.
\end{equation*}
Now we employ free boundary condition \eqref{b2} and note that the symmetry of the Cauchy tensor leads {to} the identity
$\boldsymbol{\sigma} : \nabla(\mathbf{v-u})=\boldsymbol{\sigma} : (\mathbf{Dv-Du})$. This brings us to the equality
\begin{equation}\label{eq_ident}
\int_{\Omega(t)}\left\{\rho(\frac{\partial \mathbf{u}}{\partial t}+(\mathbf{u}\cdot\nabla \mathbf{u}))(\mathbf{\mathbf{v-u}})
+ \boldsymbol{\sigma} : (\mathbf{Dv-Du}) -\blf\cdot(\mathbf{\mathbf{v-u}})\right\}\dx=
-\int_{\Gamma(t)}\gamma \kappa \mathbf{n}\cdot(\mathbf{v-u})\,\ds.
\end{equation}
As the next step, we decompose the stress tensor into deviatoric and volumetric parts:
$\boldsymbol{\sigma}=\boldsymbol{\tau}-p\bI$ (the decomposition is formal in the plug region).
We treat the stress term in \eqref{eq_ident} separately in the flow and plug regions of $\Omega(t)$.
For the flow region $\Omega_f(t)$, we employ the first constitutive relation from \eqref{eq_const} and further
apply the Cauchy-Schwarz inequality $\mathbf{Du}: \mathbf{Dv} \le |\mathbf{Du}||\mathbf{Dv}|$. We get
\begin{multline}\label{tau part1}
\int_{\Omega_f(t)} \boldsymbol{\sigma}: (\mathbf{Dv-Du})\,\dx=\int_{\Omega_f(t)}(K|\mathbf{Du}|^{\alpha-1}
+\tau_s |\mathbf{Du}|^{-1})\mathbf{Du}:(\mathbf{Dv-Du})\,\dx
\\ =
\int_{\Omega_f(t)}\left\{K|\mathbf{Du}|^{\alpha-1} \mathbf{Du} : (\mathbf{Dv}-\mathbf{Du})+
 \tau_s\left(\, |\mathbf{Du}|^{-1}(\mathbf{Du}: \mathbf{Dv})-|\mathbf{Du}|\right)\right\}\dx
\\ \le
\int_{\Omega_f(t)}\left\{K|\mathbf{Du}|^{\alpha-1} \mathbf{Du} : (\mathbf{Dv}-\mathbf{Du})+ \tau_s (|\mathbf{Dv}|-|\mathbf{Du}|)\right\}\dx.
\end{multline}
The pressure term disappears above since both $\bv$ and $\bu$ are divergence free. The same arguments and
the second constitutive relation from \eqref{eq_const} give for the plug region $\Omega_p(t)=\Omega(t)\setminus
\overline{\Omega}_f(t)$:
\begin{multline}\label{tau part2}
\int_{\Omega_p(t)} \boldsymbol{\sigma}: (\mathbf{Dv-Du})\,\dx=\int_{\Omega_p(t)} \boldsymbol{\tau}: \mathbf{Dv}\,\dx
\le
\sup_{\Omega_p(t)}|\boldsymbol{\tau}|\int_{\Omega_p(t)}|\mathbf{Dv}|\,\dx \le
\tau_s\int_{\Omega_p(t)}|\mathbf{Dv}|\,\dx \\{\footnotesize (\text{since}~~|\mathbf{Du}|=0)}=
\int_{\Omega_p(t)}\left\{K|\mathbf{Du}|^{\alpha-1} \mathbf{Du} : (\mathbf{Dv}-\mathbf{Du})+
\tau_s (|\mathbf{Dv}|-|\mathbf{Du}|)\right\}\dx.
\end{multline}
Substituting  \eqref{tau part1} and \eqref{tau part2} back into \eqref{eq_ident} gives the inequality
\begin{multline}\label{inequality}
\int_{\Omega(t)}\left\{\rho(\frac{\partial \mathbf{u}}{\partial t}+\mathbf{u}\cdot\nabla \mathbf{u})   \cdot(\mathbf{v-u})
+K|\mathbf{Du}|^{\alpha-1} \mathbf{Du} : (\mathbf{Dv}-\mathbf{Du})+ \tau_s (|\mathbf{Dv}|-|\mathbf{Du}|)\right\}\dx \\
-\int_{\Omega(t)}\blf\cdot(\mathbf{v-u})\,\dx +\int_{\Gamma(t)}  \gamma \kappa \mathbf{n}\cdot(\mathbf{v-u})\,\ds\ge 0.
\end{multline}

The arguments in this section are valid if a solution $\bu$ is sufficiently smooth. The sufficient regularity
assumptions would be $\bu\in W^{1,\alpha+1}(\Omega(t))$, $\frac{\partial \mathbf{u}}{\partial t}\in L^2(\Omega(t))$, and $\Omega(t)$ is bounded and  has $C^2$ boundary  for almost  all $t>0$. {For the case of general boundary condition on the normal stress tensor, inequality \eqref{inequality} is found in \cite{huilgol2002variational}.}

We summarize the result of this section: \textit{A sufficiently smooth solution $\bu$ to
\eqref{N-S}--\eqref{b2} satisfies the variational inequality \eqref{inequality} for almost all $t>0$ and for any $\bv\in H^1(\Omega(t))$ such that
$\Div\bv=0$.}

\section{Energy balance} \label{s_energy}
The energy balance for the solution to the free-surface flow problem \eqref{N-S}--\eqref{b2} follows from the
variational inequality \eqref{inequality}. To show this, we first recall a few helpful identities.
We shall assume that $\Gamma(t)$ is sufficiently smooth and closed for all $t \in [0,T]$.
For a smooth function $g$ defined on $\bigcup\limits_{t \in [0,T]} \Omega(t)\times \{t\}$, the Reynolds transport theorem
gives the relation
\begin{equation}\label{reynold}
\frac{d}{dt} \int_{\Omega(t)} g\, \dx = \int _{\Omega(t)} \frac{\partial g}{\partial t} \dx + \int_{\Gamma(t)} v_{\Gamma} g\, \ds.
\end{equation}
Thanks to the kinematic condition \eqref{eq_vu} on the normal velocity of $\Gamma$
and  $ \mbox{div}\mathbf{u}=0$, \eqref{reynold} yields the identity
\begin{equation}
\frac{d}{dt} \int_{\Omega(t)} g \,\dx=\int_{\Omega(t)}\left(\frac{\partial g}{\partial t}+(\mathbf{u} \cdot \nabla) g\right) \dx. 
\end{equation}

Recall the definition of the surface gradient and divergence operators: $\nabla_{\Gamma} q = \nabla q - (\bn \cdot \nabla q)\bn$
and  $  \mbox{div}_{\Gamma} \mathbf{g}= \mbox{tr}(\nabla _{\Gamma}\mathbf{g})$,
 which are the intrinsic surface quantities and do not depend on extensions of a scalar function $q$
 and a vector function $\mathbf{g}$ off the surface. The integration by parts formula over a closed smooth
 surface $\Gamma$ reads
 \begin{equation}\label{int_by_parts}
 \int_{\Gamma} \left(\,q(\mbox{div}_{\Gamma}\mathbf{g})+ \mathbf{g} \cdot \nabla_{\Gamma}q\,\right)\ds= \int_{\Gamma} \kappa (\mathbf{g} \cdot \mathbf{n})q\, {\rm d}\bs,
 \end{equation}
 where $\kappa$ denotes the (doubled) surface mean curvature as in \eqref{inequality}.
Finally, for $\Gamma(t)$ passively advected by a flow field $\mathbf{u}$, the Leibniz formula
gives
 \begin{equation}\label{ll2}
 \frac{d}{dt} \int_{\Gamma(t)} g \, \ds= \int_{\Gamma(t)} \left(\frac{\partial g}{\partial t}
 +(\mathbf{u} \cdot \nabla) g+ g\, \mbox{div}_{\Gamma} \mathbf{u}\right)\, \ds.
 \end{equation}
\medskip

 Now we are prepared to deduce the problem energy balance from \eqref{inequality}. As the first step, we test
  \eqref{inequality} with $\bv=0$ and $\bv=2\bu$. Comparing two resulting inequalities, we obtain the equality
\[
\int_{\Omega(t)}\left\{\rho(\frac{\partial \mathbf{u}}{\partial t}\cdot\mathbf{u}+(\mathbf{u}\cdot \nabla \mathbf{u})\cdot\mathbf{u}) + K|\mathbf{Du}|^{\alpha+1}+\tau_s |\mathbf{Du}|\right\}\dx
+\int_{\Gamma(t)}  \gamma \kappa \mathbf{n}\cdot\mathbf{u}\,\ds= \int_{\Omega(t)}\blf\cdot\bu\,\dx.
\]
We rewrite the first two terms as $\frac12\int_{\Omega(t)}\rho(\frac{\partial |\mathbf{u}|^2}{\partial t}
+\mathbf{u}\cdot \nabla |\mathbf{u}|^2)\,\dx$ and apply  the Reynolds transport formula. This gives the identity
 \begin{align}\label{ni}
 \frac{d}{dt}\int_{\Omega(t)}\frac{\rho|\mathbf{u}|^2}{2}\dx + \int_{\Omega(t)}\left(K|\mathbf{Du}|^{\alpha+1}+\tau_s |\mathbf{Du}|\right)\dx
+\int_{\Gamma(t)}  \gamma \kappa \mathbf{n}\cdot\mathbf{u}\,\ds=  \int_{\Omega(t)}\blf\cdot\bu\,\dx.
\end{align}

With the help of integration by parts \eqref{int_by_parts} over $\Gamma=\Gamma(t)$
and the Leibniz formula we calculate:
\begin{equation*}
 \int_{\Gamma(t)} \kappa \mathbf{(n \cdot u)}\ds=  \int_{\Gamma(t)} \mbox{div}_{\Gamma} \mathbf{u}\ds =\frac{d}{dt}  \int_{\Gamma(t)} 1 \ds = \frac{d}{dt} |\Gamma(t)|,
\end{equation*}
where $|\Gamma(t)|$ denotes the area of the free surface.
Employing these relations in  \eqref{ni} leads to the following \textit{energy balance for the solution of \eqref{N-S}--\eqref{b2}}:
 \begin{equation}\label{balance}
\frac{d}{dt}\left(\int_{\Omega(t)}\frac{\rho|\mathbf{u}|^2}{2}\dx +\gamma |\Gamma(t)|\right) +
\int_{\Omega(t)}\left(K|\mathbf{Du}|^{\alpha+1}+\tau_s |\mathbf{Du}|\right)\dx
=  \int_{\Omega(t)}\blf\cdot\bu\,\dx.
\end{equation}
The energy balance \eqref{balance} has the form
\[
\frac{d}{dt} E_{\rm total}(t) = -D(t)+ W_{\rm ext}(t),
\]
where the total energy $E_{\rm total}(t)$ is the sum of kinetic energy
$E_{\rm kin}(t)=\int_{\Omega(t)}\frac{\rho|\mathbf{u}|^2}{2}\dx$
and potential energy $E_{\rm free}(t)=\gamma |\Gamma(t)|+{\rm const}$. The rate of change of $E_{\rm total}$ is
balanced by the internal energy dissipation
\begin{equation}\label{diffusion}
D(t)=\int_{\Omega(t)}\left(K|\mathbf{Du}|^{\alpha+1}+\tau_s |\mathbf{Du}|\right)\dx
\end{equation}
and the work of external forces
\[
W_{\rm ext}(t)=\int_{\Omega(t)}\blf\cdot\bu\,\dx.
\]

\begin{remark}\rm \label{rem1}
Since there is no explicit dissipation mechanism for the free surface energy $E_{\rm free}(t)$ in
\eqref{balance}, it is not easy to obtain directly from \eqref{balance} \textit{a priori} estimates for the solution which would be sufficient for showing the (local) well-posedness of the problem.
Solonnikov in \cite{Solonnikov91} was {the} first to study the solvability of the Newtonian fluid
free-surface flow problem subject to surface tension forces. His proof does not directly rely on energy estimates, but rather on Fourier-Laplace transform techniques, which required the use of exponentially weighted anisotropic Sobolev--Slobodeskii spaces with fractional-order spatial derivatives.
Further, energy methods to establish  new space-time estimates for the Newtonian flows were developed in
\cite{CS2002} and semigroup approach to establish the existence  was used in \cite{schweizer1997free}.
{None} of these analyses are known to extend to viscoplastic fluid flow problems with free surfaces and surface tension forces. { If one is interested in the existence of the arrested state or the finite stopping time, then the available analysis requires a lower bound for the plastic dissipation term of the form $\big(\int_{\Omega(t)}|\mathbf{u}|^2\dx\big)^{\frac12}\le C_{bd}(t)\int_{\Omega(t)}\tau_s |\mathbf{Du}|\dx$. The bound is feasible for certain one-dimensional flows and for the flow in a long pipe of a constant cross-section \cite{glowinski2008lectures,huilgol2002variational,TemamStrang,Kohn82}. However, in a more general case of $\Omega\in\mathbb{R}^d$, the estimate implies the embedding $W^{1,1}(\Omega)\hookrightarrow L^2(\Omega)$ , which is known to be valid only for $d\le2$. We note that this fundamental difficulty arises within the existing framework regardless of the form of exterior forces and also for the fixed (time-independent) domain.} For free-surface flows, one also needs to control the constant $C_{bd}(t)$ for all possible (\textit{a priori} unknown) shapes of $\Omega(t)$.
\end{remark}

For the reasons outlined in Remark~\ref{rem1} above, it is not clear what can be concluded about the energy decay or the existence of a finite cessation time for  the problem  \eqref{N-S}--\eqref{b2}  solely from the energy balance \eqref{balance}. 
Further in this paper, we look {closely} at the problem of the existence of $T_f$ using the example of { a} freely oscillating viscoplastic droplet rather than {considering} a general flow solving  \eqref{N-S}--\eqref{b2}. The classical problem of oscillating droplet of viscous  incompressible fluid with surface tension forces was  treated by Lamb in \cite{Lamb}. { Lamb assumed an irrotational velocity field and used the dissipation method to evaluate the effect of the viscosity on the decay of the oscillations.
An exact solution of this problem for the Newtonian case is found in the analysis by Miller and Scriven \cite{miller1968oscillations} of the oscillations of a fluid droplet immersed in another fluid.
The viscous effects on the perturbed spherical flows were further studied in \cite{prosperetti1980free}. Those studies indicated that the no-slip condition on the interface between two fluids is a major source of vorticity production in the problem, while the irrotational velocity field is an adequate approximation in the \textit{viscous} case, if the interface is free  and one of two  fluids is a gas of negligible density and viscosity. In the present study, the exterior is vacuum and 
we enforce no condition on tangential velocities. Hence for the analysis we accept the irrotational velocity field assumption.
In section~\ref{s_num} we include the results of a few numerical experiments, which illustrate the plausibility of this assumption. For the extended discussion of the plausibility of the vorticity-free approximation for the oscillating viscous droplet problem
we refer to \cite{joseph2006potential,padrino2008purely}.
In the next section, we shall see that for the vorticity-free approximation,   the energy  balance \eqref{balance} yields the existence of a finite cessation time $T_{f}$.}  

\section{Free oscillations of a viscoplastic drop}\label{s_drop}

In this section, we study  free oscillations of a viscoplastic droplet near its equilibrium state.
Assuming rotational symmetry\footnote{The rotational symmetry is assumed for the {clarity of presentation}. The arguments can be extended to more general perturbations.},  the initial shape of the droplet is given by a perturbation of the sphere
\begin{equation}\label{initial}
r= r_0(1+\tilde{\ep} \sum_{n\ge1}c_n\mathcal{H}_n(\theta,\varphi)),
\end{equation}
where $(r,\theta,\varphi)$ are spherical coordinates,  $\mathcal{H}_n$, $n=1,2,\dots$, is the $n$th spherical harmonic and $\tilde{\ep}$ is small,  {$\tilde{\ep}\ll1$.} We denote by $S_0$ the unperturbed sphere of radius $r_0$ and without loss of
generality assume that $\mathcal{H}_n$ are normalized, i.e. $\|\mathcal{H}_n\|_{L^2(S_0)}=1$, and $\sum_{n\ge1}c_n^2=1$.
The fluid is assumed to be at rest at time $t =0$ and  $\bf{f} = 0$ for all $t\ge0$.
At $t=0$ the mean curvature of the surface is not constant, and an unbalanced surface tension force causes {the}
droplet oscillation. Following \cite{Lamb}, we consider the evolution of the droplet surface given by
\begin{equation}\label{osc}
r= r_0+\sum_{n\ge1} A_n(t)\mathcal{H}_n(\theta,\varphi)=:r_0+\sum_{n\ge1}\xi_n.
\end{equation}
In the absence of dissipation, Lamb showed that $A_n=r_0c_n\tilde{\ep}\sin(\sigma_n t+\alpha_n)$, where the period of oscillations depends on surface tension, fluid density, the harmonic's index $n$, and $r_0$. 
Our plan for the {analysis}  is the following: For the droplet evolving according to   \eqref{osc}  we find the velocity potential and compute  $E_{\rm free}$,  $E_{\rm kin}$,  and the viscous energy dissipation from \eqref{balance} in terms of $A_n(t)$ and its derivatives. Examining the resulting system of ODEs for $A_n(t)$ we recover the classical results of Lamb about the period and decay of oscillations  for the Newtonian droplet. This result is {of further} help when we treat the yield stress case.  The viscoplastic dissipation in \eqref{balance} is  estimated from below. This provides us with a differential inequality for $A_n(t)$ for all $t>0$. Analysis of this differential inequality yields the existence of a finite stopping time,
i.e. $A_n(t)=0$ for all $t\ge T_f$ and $n\ge1$. 
\medskip

The velocity potential $\phi$ of irrotational flow of incompressible fluid is a harmonic function for all $t>0$.
We seek $\phi$ in the form of volume spherical harmonics series
\[
\phi = \sum_{n\ge1} B_n(t) \frac{r^n}{r_0^n} \mathcal{H}_n
\]
Let $\xi:=\sum_{n\ge1}\xi_n$. The kinematic boundary condition \eqref{eq_vu} can be written as
\begin{equation}\label{b2a}
\frac{\partial\phi}{\partial r} =  \frac{\partial\xi }{\partial t}\quad\text{on}~~S_0.
\end{equation}
This gives
\begin{equation}\label{ba}
\frac{n}{r_0}B_n(t)=\frac{dA_n(t)}{dt}.
\end{equation}
{Further in the paper, we always assume that $A_n$ and $B_n$  are functions of time. In most instances, we shall write $A_n$ and $B_n$ instead of $A_n(t)$ and $B_n$(t).}

With the help of $\Delta\phi=0$ and dropping higher order { with respect to $\tilde{\ep}$} terms one computes the kinetic energy:
\begin{equation}\label{E_kin}
E_{\rm kin}(t)=\frac\rho2\int_{\Omega(t)}|\nabla\phi|^2\dx=\frac\rho2\int_{\Gamma(t)}\phi \frac{\partial\phi}{\partial \bn} \ds\simeq
\frac\rho2\int_{S_0}\phi \frac{\partial\phi}{\partial r} \ds=\frac\rho{2r_0}\sum_{n\ge1}nB^2_n\int_{S_0}\mathcal{H}_n^2 \ds.
\end{equation}
Recall that $\int_{S_0}\mathcal{H}_n\mathcal{H}_m\ds =\delta_m^n$. Employing \eqref{ba}, we find the {rate of change} of $E_{\rm kin}$:
\begin{equation}\label{E_kin_var}
\frac{d}{dt}E_{\rm kin}(t)=\frac\rho{r_0}\sum_{n\ge1}n\frac{dB_n}{dt} B_n =\rho\sum_{n\ge1} \frac{r_0}{n}\frac{d^2A_n}{dt^2}\frac{dA_n}{dt}.
\end{equation}
 For the potential energy, we first calculate
\[
\begin{aligned}
\kappa=\frac1{R_1}+\frac1{R_2}&=\frac{2}{r_0}-\frac{2\xi}{r_0^2}-\frac{1}{r_0^2}\left\{\frac1{\sin^2\theta}\frac{\partial^2\xi}{\partial \phi^2}+\frac1{\sin\theta}\frac{\partial}{\partial \theta}\left({\sin\theta}\frac{\partial\xi}{\partial \theta}\right)\right\}\\&=\frac{2}{r_0}-\sum_{n\ge1}\left\{\frac{2A_n\mathcal{H}_n}{r_0^2}-\frac{n(n+1)}{r_0^2} A_n\mathcal{H}_n\right\}
=\frac{2}{r_0}+\sum_{n\ge1}\frac{A_n}{r_0^2}(n(n+1)-2)\mathcal{H}_n.
\end{aligned}
\]
where ${R}_{1},{R}_{2}$ are the radii  of principal of curvature of the surface. Noting $\int_{S_0}\mathcal{H}_n=0$, $n\ge1$, we get for the {rate of change} of the potential energy
\begin{equation*}
\begin{aligned}
\frac{d}{dt}E_{\rm free}(t)&=\gamma\int_{\Gamma(t)} \kappa \mathbf{(n \cdot u)}\ds\simeq \gamma\int_{S_0}\kappa \frac{\partial\phi}{\partial r} \ds=\sum_{n\ge1}\frac{\gamma(n-1)(n+2)n}{r_0^3}B_nA_n\int_{S_0}\mathcal{H}_n^2 \ds
\\=&\sum_{n\ge1}\frac{\gamma\,(n-1)(n+2)}{r_0^2}\frac{dA_n}{dt}A_n.
\end{aligned}
\end{equation*}
This and \eqref{E_kin_var} gives the {rate of change} of the total energy
\begin{equation}\label{E_total0}
\frac{d}{dt}E_{\rm total}(t)=\sum_{n\ge1}\left\{ \frac{r_0\rho}{n}\frac{d^2A_n}{dt^2}\frac{dA_n}{dt}+\frac{\gamma\,(n-1)(n+2)}{r_0^2}\frac{dA_n}{dt}A_n\right\}.
\end{equation}
For the \textit{ideal} fluid setting $\frac{d}{dt}E_{\rm total}(t)=0$ one finds $A_n(t)=\widehat{A}_0\sin(\sigma_n t+\alpha_n)$ with the frequency
\begin{equation}\label{E_total}
{\sigma_n}=\frac{\sqrt{\gamma\, n(n-1)(n+2)}}{r_0^\frac32\rho^{\frac12}}
\end{equation}
as in \cite{Lamb}. For the Bingham fluid, one should account for viscous and plastic dissipation.
Thanks to $\nabla\cdot\bu=0$ and $\nabla\times\bu=0$, we have
\[
\int_{\Omega(t)}|\mathbf{D}\bu|^2\dx=
\int_{\Omega(t)}|\nabla\bu|^2\dx=\int_{\Gamma(t)}\bu\cdot\frac{\partial\bu}{\partial\bn}\ds
=\frac12\int_{\Gamma(t)}\frac{\partial|\bu|^2}{\partial\bn}\ds\simeq \frac12\int_{S_0}\frac{\partial|\nabla\phi|^2}{\partial r}\ds.
\]
Note that $\frac{\partial\phi_n}{\partial r}=\frac{n}{r}\phi_n$, with $\phi_n=B_n \frac{r^n}{r_0^n} \mathcal{H}_n,$ and for the Laplace--Beltrami operator on a sphere of radius $r$, it holds $\Delta_\Gamma\phi_n=-n(n+1)r^{-2}\phi_n$.
With the help of these identities and the surface integration  by parts formula \eqref{int_by_parts}, we handle the integral on
the righthand side as follows:
\[
\begin{aligned}
\int_{S_0}\frac{\partial|\nabla\phi|^2}{\partial r}\ds&=\int_{S_0}\frac{\partial}{\partial r}\left(|\nabla_\Gamma\phi|^2+ \left|\frac{\partial\phi}{\partial r}\right|^2\right)\ds\\&
=\int_{S_0}\frac{\partial}{\partial r}\left(-\phi \Delta_\Gamma\phi+ \left|\frac{\partial\phi}{\partial r}\right|^2\right)\ds\\
&=\sum_{n\ge1}\int_{S_0}\frac{\partial}{\partial r}\left(n(n+1)r^{-2}\phi^2_n+ n^2r^{-2}\phi^2_n\right)\ds.
\end{aligned}
\]
Substituting $\phi_n=B_n(r/r_0)^n \mathcal{H}_n$, we find
\begin{equation}\label{dissip}
\begin{aligned}
\int_{\Omega(t)}|\mathbf{D}\bu|^2\dx&=\sum_{n\ge1} n(n-1)(2n+1)r_0^{-3}[B_n]^2\int_{S_0}\mathcal{H}_n^2 \ds\\&
=\sum_{n\ge1}\frac{(n-1)(2n+1)}{nr_0}\left|\frac{dA_n}{dt}\right|^2.
\end{aligned}
\end{equation}
Thus, for the \textit{Newtonian} fluid, we get from \eqref{balance}, \eqref{E_total0} and \eqref{dissip} that $A_n$ satisfy
\begin{equation}\label{Anewton}
\frac{d^2A_n}{dt^2}+\frac{K\,(n-1)(2n+1)}{r_0^{2}\rho}\frac{dA_n}{dt}+ \frac{\gamma\,n(n-1)(n+2)}{r_0^{3}\rho}A_n =0,\quad \text{for}~~n=2,3,\dots.
\end{equation}
When the determinant of the  characteristic equation for some $n>1$ is non-negative (viscosity dominates over surface tension),
then the corresponding harmonic does not contribute to oscillations and, using the initial condition $\frac{dA_n}{dt}=0$ (since the fluid is assumed at rest at initial time, i.e. $\phi=0$),  one finds
\begin{equation}\label{decay}
A_n(t)=A_n(0)\frac{\lambda_n^1\exp(\lambda_n^2 t)-\lambda_n^2\exp(\lambda_n^1 t)}{\lambda_n^1-\lambda_n^2},
\end{equation}
where $\lambda_n^{1,2}<0$ are corresponding {real} eigenvalues.
{For complex eigenvalues, we observe oscillatory behavior.}
The amplitude of the oscillations for $n$th harmonic decays exponentially:
\[
A_n(t)=\widehat{A}_n(0)\exp(-d_n\,t)\sin(\sigma_nt+\alpha_n),\quad\text{with}~~d_n=-Re(\lambda_n^1)=\frac{K\,(n-1)(2n+1)}{2r_0^{2}\rho}
\]
and $\sigma_n$ from \eqref{E_total}, $\widehat{A}_n(0)=A_n(0)/\sin\alpha_n$.
We note that for any fixed positive problem parameters $r_0$, $\rho$, $\gamma$, $K$, it holds
\begin{equation}\label{n_min}
|A_n| \le A(0)\exp(-c_d n^2\,t), 
\end{equation}
with a constant $c_d$ depending only on the parameters of the problem.

Further we  consider the effect of the plastic dissipation. To this end, we need the following trace inequality for functions of
bounded variation in an $N$-dimensional ball~\cite{cianchi2012sharp}:
\[
\|u-|\partial B|^{-1}\int_{\partial B}u\ds\|_{L^1(\partial B)}\le \frac{N\sqrt{\pi}\,\Gamma(\frac12(N+1))}{2\Gamma(\frac12(N+2))}
\|\nabla u\|_{L^1(B)}~\overset{N=3}{=}~2\|\nabla u\|_{L^1(B)}.
\]
Noting that for  irrotational flow $\mathbf{D}\bu=\nabla\bu$ and due to axial symmetries $\int_{S_0}u_i=0$, $i=1,2,3$, we apply the above inequality componentwise and we  estimate  the plastic dissipation to be \textit{at least}
\begin{equation}\label{dissip2.1}
\begin{aligned}
\tau_s\int_{\Omega(t)}|\mathbf{D}\bu|\dx&=
\tau_s \int_{\Omega(t)}|\nabla\bu|\dx\simeq \tau_s \int_{\Omega_0}|\nabla\bu|\dx\ge \frac{\tau_s}3 \int_{\Omega_0}|\nabla\bu|_{\ell^1}\dx\\ &\ge \frac{\tau_s}{6}\int_{S_0}|\bu|_{\ell^1}\ds
=\frac{\tau_s}{6}\int_{S_0}|\nabla\phi|_{\ell^1}\ds\ge \frac{\tau_s}{6}\int_{S_0}|\nabla\phi|_{\ell^2}\ds\\ &
=\frac{\tau_s}{6}\int_{S_0}\left(|\nabla_\Gamma \phi|^2+\left|\frac{\partial\phi}{\partial r}\right|^2\right)^{\frac12}\ds.
\end{aligned}
\end{equation}
Let $C_{\rm emb}$ be optimal constant from the following Sobolev  embedding inequality for the two-dimensional sphere $S_0$:
\[
\|u\|_{L^2(S_0)}\le C_{\rm emb}\|u\|_{W^{1,1}(S_0)}\quad\text{for}~u\in W^{1,1}(S_0),~~\text{s.t.} \int_{S_0}u=0.
\]
Applying this result, we proceed with the estimate on plastic dissipation from below as follows.
\begin{equation}\label{dissip2}
\begin{aligned}
\tau_s\int_{\Omega(t)}|\mathbf{D}\bu|\dx&\ge \frac{\tau_s}{6}\int_{S_0}|\nabla_\Gamma \phi|\ds
\ge\frac{\tau_s}{6 C_{\rm emb}}\left(\int_{S_0}|\phi|^2\ds\right)^{\frac12}\\ & =\frac{\tau_s}{6 C_{\rm emb}}\left(\int_{S_0}|\sum_{n\ge1} B_n \mathcal{H}_n|^2\ds\right)^{\frac12} =\frac{\tau_s}{6 C_{\rm emb}}\left(\sum_{n\ge1} B_n^2\right)^{\frac12}\\&=\frac{\tau_s r_0}{6 C_{\rm emb}}\left(\sum_{n\ge1}n^{-2}\left|\frac{dA_n}{dt}\right|^2\right)^{\frac12}.
\end{aligned}
\end{equation}

\subsection{Finite stopping time for Bingham drop}
We first treat the case of the flow index $\alpha=1$ (Bingham fluid).
From \eqref{dissip} and \eqref{dissip2} one gets the lower bound for the
total internal energy dissipation \eqref{diffusion} of the viscoplastic droplet
\begin{multline}\label{diffusion_bound}
D(t)=\int_{\Omega(t)}\left(K|\mathbf{Du}|^{2}+\tau_s |\mathbf{Du}|\right)\dx \\ \ge
\sum_{n\ge1}\frac{K(n-1)(2n+1)}{nr_0}\left|\frac{dA_n}{dt}\right|^2 + \frac{\tau_s r_0}{6 C_{\rm emb}}\left(\sum_{n\ge1}n^{-2}\left|\frac{dA_n}{dt}\right|^2\right)^{\frac12}
\end{multline}
Substituting this estimate to the total energy balance relation one obtains  the following differential
inequalities for $A_n$:
\begin{multline}\label{eq_At}
\sum_{n\ge1}\left\{\frac{\rho}{2n} \frac{d}{dt}\left|\frac{dA_n}{dt}\right|^2+\frac{K\,(n-1)(2n+1)}{n\,r_0^{2}}\left|\frac{dA_n}{dt}\right|^2+ \frac{\gamma\,(n-1)(n+2)}{2r_0^{3}}\frac{d |A_n|^2}{dt}\right\}\\  + \frac{\tau_s }{6\pi C_{\rm emb}}\left|\sum_{n\ge1}n^{-2}\left|\frac{dA_n}{dt}\right|^2\right|^{\frac12} \le 0.
\end{multline}
\medskip

Based on \eqref{eq_At} and the previous analysis, we show that there exists such finite $T_f$ that $A_n=0$ for all $n\ge1$ and $t>T_f$. To this end, we first estimate
the third (surface tension) term with the help of the Cauchy inequality:
\begin{equation}\label{aux1}
\begin{aligned}
\sum_{n\ge1}\frac{\gamma\,(n-1)(n+2)}{2r_0^{3}}\frac{d |A_n|^2}{dt} &= \sum_{n\ge1}\frac{\gamma\,(n-1)(n+2)}{r_0^{3}}\frac{d A_n}{dt}A_n\\
&\le \frac{\gamma\,}{r_0^{3}}\left(\sum_{n\ge1}(n-1)^2(n+2)^2n^2A_n^2\right)^{\frac12}\left(\sum_{n\ge1}n^{-2}\left|\frac{d A_n}{dt}\right|^2\right)^{\frac12}.
\end{aligned}
\end{equation}
From the study of {purely} viscous case, when there is no additional plastic dissipation, we know that $A_n$ decay at least exponentially with the decay factors not less than $-c_d n^2$, see \eqref{n_min}. If we
\textit{assume} that adding the plastic dissipation can only contribute to the energy decay in a given
harmonic,  we conclude that there exists such finite time $T_1$ that
\begin{equation}\label{aux2}
\frac{\gamma\,}{r_0^{3}}\left(\sum_{n\ge1}(n-1)^2(n+2)^2n^2A_n^2\right)^{\frac12}\le \frac{\tau_s }{12 C_{\rm emb}}\quad\text{for}~t\ge T_1.
\end{equation}
Using this and \eqref{aux1} in \eqref{eq_At}, we get
\begin{equation}\label{eq_Bt}
\sum_{n\ge1}\left\{\frac{\rho n}{2} \frac{d}{dt}B_n^2+\frac{K\,(n-1)(2n+1)n}{r_0^{2}}B_n^2\right\}+  \frac{\tau_sr_0 }{12 C_{\rm emb}}\left|\sum_{n\ge1}B_n^2\right|^{\frac12} \le 0\quad\text{for}~t\ge T_1.
\end{equation}
For the sake of convenient notation we also {make} the substitution  $\frac{dA_n}{dt}=\frac{n}{r_0}B_n$.  Further we use the H\"{o}lder inequality to estimate the plastic dissipation term from below:
\[
\sum_{n\ge1} n B_n^2\le  \left(\sum_{n\ge1} n^p B_n^{(2-\alpha)p}\right)^{\frac1p}
\left(\sum_{n\ge1}  B_n^{\alpha q}\right)^{\frac1q}~\overset{p=\frac32,\,q=3,\,\alpha=\frac23}{=}
\left(\sum_{n\ge1} n^{\frac32} B_n^2\right)^{\frac23} \left(\sum_{n\ge1}  B_n^{2}\right)^{\frac13}.
\]
Thanks to the Young inequality we have
\begin{equation*}
\left(\sum_{n\ge1} n B_n^2\right)^{\frac34}\le  \frac1{2\delta} \left(\sum_{n\ge1}  B_n^{2}\right)^{\frac12}+\frac\delta2\sum_{n\ge1} n^{\frac32} B_n^{2}
\quad\forall~\delta>0,
\end{equation*}
or after obvious rearrangement of terms
\begin{equation}\label{aux3b}
2\delta\left(\sum_{n\ge1} n B_n^2\right)^{\frac34}- \delta^2\sum_{n\ge1} n^{\frac32} B_n^2\le  \left(\sum_{n\ge1}  B_n^2\right)^{\frac12}
\quad\forall~\delta>0,
\end{equation}
Thanks to the  kinetic energy decay, we may always assume that $T_1$ is such that $E_{\rm kin}(t)\le\frac\rho2$ for $t\ge T_1$ and so $B_1^2\le \frac2\rho E_{\rm kin}\le1$ for $t\ge T_1$.
Hence for $\delta\in(0,1]$ it holds $\delta^2B_1^2\le \delta\left(\sum_{n\ge1} n B_n^2\right)^{\frac34}$. Now \eqref{aux3b} yields
\begin{equation}\label{aux3a}
\delta\left(\sum_{n\ge1} n B_n^2\right)^{\frac34}- \delta^2\sum_{n\ge2} n^{\frac32} B_n^2\le  \left(\sum_{n\ge1}  B_n^2\right)^{\frac12}
\quad\forall~\delta>0,
\end{equation}
If we substitute \eqref{aux3a}   in \eqref{eq_Bt} with  $\delta$ satisfying
\begin{equation}\label{delta1}
0<\delta^2\le\frac{5K}{r_0^2\sqrt{2}},
\end{equation}
then $\frac12$ of the viscous term kills the negative term on the left hand side of \eqref{aux3a}.
Further, for the viscous term in \eqref{eq_Bt}, the {following holds trivially}
\begin{equation}\label{aux3}
\sum_{n\ge1}\frac{K\,(n-1)(2n+1)n}{2r_0^{2}}B_n^2\ge \frac{5K }{2r_0^2}\sum_{n\ge2} n B_n^2.
\end{equation}
Finally, we get control of $ \frac{5K }{2r_0^2}B_1^2$ with the help of the viscoplastic term.
Again, thanks to the  kinetic energy decay, we may  assume that $T_1$ is sufficiently large such that  for $t\ge T_1$ the
coefficient $B_1$ is small to satisfy the inequality
\begin{equation*}
\frac{5K }{2r_0^2}B_1^2\le \frac{\tau_s r_0\delta}{24 C_{\rm emb}}\left(\sum_{n\ge1} n B_n^2\right)^{\frac34}.
\end{equation*}
Thus, using \eqref{aux3a}, \eqref{aux3}  in \eqref{eq_Bt} and choosing $\delta$ satisfying \eqref{delta1},  we arrive at the following differential inequality for the quantity $\widehat{B}:=\sum_{n\ge1} n B_n^2$:
\[
\frac{\rho}{2}\frac{d\widehat{B}}{dt}+\frac{5K }{2r_0^2}\widehat{B} +\frac{\tau_s r_0\delta}{24 C_{\rm emb}}\widehat{B}^{\frac34}\le0\quad\text{for}~~t\ge T_1.
\]
The ODE $y'+ c_1y+c_2y^s=0$  is solved by $y^{1-s}=(y^{1-s}(0)+c_2c_1^{-1})e^{-(1-s)c_1t}-c_2c_1^{-1}$ for $t\ge0$, $s\neq 1$.
Hence, the comparison theorem provides us with the bound
\[
\widehat{B}^{\frac14}\le (\widehat{B}^{\frac14}(T_1)+c_2c_1^{-1})e^{-\frac{c_1(t-T_1)}4}-c_2c_1^{-1},\quad \text{for all}~t\ge T_1,
\]
with $c_1=\frac{5K }{r_0^2\rho}$, $c_2=\frac{\tau_s r_0\delta}{12\rho C_{\rm emb}}$.
We conclude that $\widehat{B}=0$ for $t\ge T_f$, with a finite stopping time $T_f$.

\begin{remark}\label{rem2}\rm
The analysis above can be simplified for $d=2$, i.e. for the problem of 2D oscillating drop. Indeed, in this case one can use the continuous embedding
$W^{1,1}(\Omega)\hookrightarrow L^2(\Omega)$, $\Omega\subset\mathbb{R}^2$, and estimate the plastic dissipation terms from below as follows (compare
to \eqref{dissip2.1}-\eqref{dissip2} and arguments below \eqref{eq_Bt}):
\[
\begin{aligned}
\tau_s\int_{\Omega(t)}|\mathbf{D}\bu|\dx&=
\tau_s \int_{\Omega(t)}|\nabla\bu|\dx\simeq \tau_s \int_{\Omega_0}|\nabla\bu|\dx\ge \widehat{C}_{\rm emb}\left(\int_{\Omega_0}|\bu|^2\dx\right)^{\frac12}\\
&=\sqrt{2}\widehat{C}_{\rm emb}E_{\rm kin}^{\frac12}(t)=\widehat{C}_{\rm emb}\left(\frac\rho{r_0}\sum_{n\ge1}nB^2_n\right)^{\frac12}.
\end{aligned}
\]
\end{remark}

\subsection{Shear thickening case} The fluid with the index {$\alpha\ge1$} fits the framework of the Bingham fluid if one notes the inequality
\[
K|\mathbf{D}\bu|^{1+\alpha}+\tau_s|\mathbf{D}\bu|\ge \min\{K,\frac{\tau_s}2\}|\mathbf{D}\bu|^{2}+\frac{\tau_s}2|\mathbf{D}\bu|\quad\text{for}~\alpha\ge1.
\]
Therefore, the above analysis applies with the viscosity coefficient $\min\{K,\frac{\tau_s}2\}$ and yield
stress $\frac{\tau_s}2$. {The analysis of the shear thinning case $\alpha<1$ is lacking at present.}

\section{Numerical experiments}\label{s_num} In this section we present the results of several numerical experiments, which illustrate the analysis of this paper. {These} experiments also study the dependence of the finite stopping time for the 3D droplet problem on various parameters. For the computer simulations we use the numerical approach developed in
\cite{nikitin2011numerical,olshanskii2013octree,nikitin2015splitting} for free-surface incompressible viscous  flows. The numerical method is built on a staggered grid finite difference octree discretization of momentum, mass conservation and level set equations. The latter is used to model the evolution of the free surface { in a bulk computational domain.}  The plasticity term is regularized by the   Bercovier-Engelman method \cite{bercovier1980finite} ($|\mathbf{D}\bu|^{-1}\to\left(|\mathbf{D}\bu|^{2}+\ep^2\right)^{-\frac12}$ in \eqref{eq_const}\,) with the regularization parameter $\ep=10^{-6}$. { We note that regularized problem may not inherit an existence of arrested state from the original problem.} However, numerical experiments demonstrate the convergent results of flow statistics for this level of values of $\ep$. { This indicates that the modelling error due to the regularization for $\ep=10^{-6}$ is minor compared to discretization errors.} The regularization allows us to overcome computational difficulties associated with the non-differentiability of the constitutive relations and hence to perform 3D computations using dynamically  refined grids towards the free surface, i.e. the refinement follows the evolution of the free surface. Such a refinement is of crucial importance for the sufficiently accurate computations of the surface tension forces. { Only those cells of the background octree mesh are active in computations, which are intersected by the surface or belong to the interior of the droplet, so no auxiliary conditions are needed on the boundary of the bulk domain.}  A second order version of the Chorin-Temam splitting method was used for time advancing and the variable time step is used subject to certain stability conditions, see details in \cite{nikitin2011numerical,nikitin2015splitting}.

\begin{table}[th]
\centering\small
\caption{\label{t0} Approximate number of total active degrees of freedom and the error in viscosity (numerical dissipation) introduced by the method for the ideal fluid.}
\label{my-label}
\begin{tabular}{r|rrrr}
\hline
 $h_{min}$& $\frac\ell{16}$  &$\frac\ell{32}$ & $\frac\ell{64}$ & $\frac\ell{128}$\\
\hline
 $\approx$ \# d.o.f.& 111333  &142405 &452681 & 1772340 \\
\hline
$\text{Error}_{\rm visc}$ & 0.0032  & 9.5750e-04    &7.2761e-04  &  4.8750e-04 \\
\hline
\end{tabular}
\end{table}

\begin{figure}[th]
 \centering
 \includegraphics[width=.49\textwidth,natwidth=1339,natheight=895]{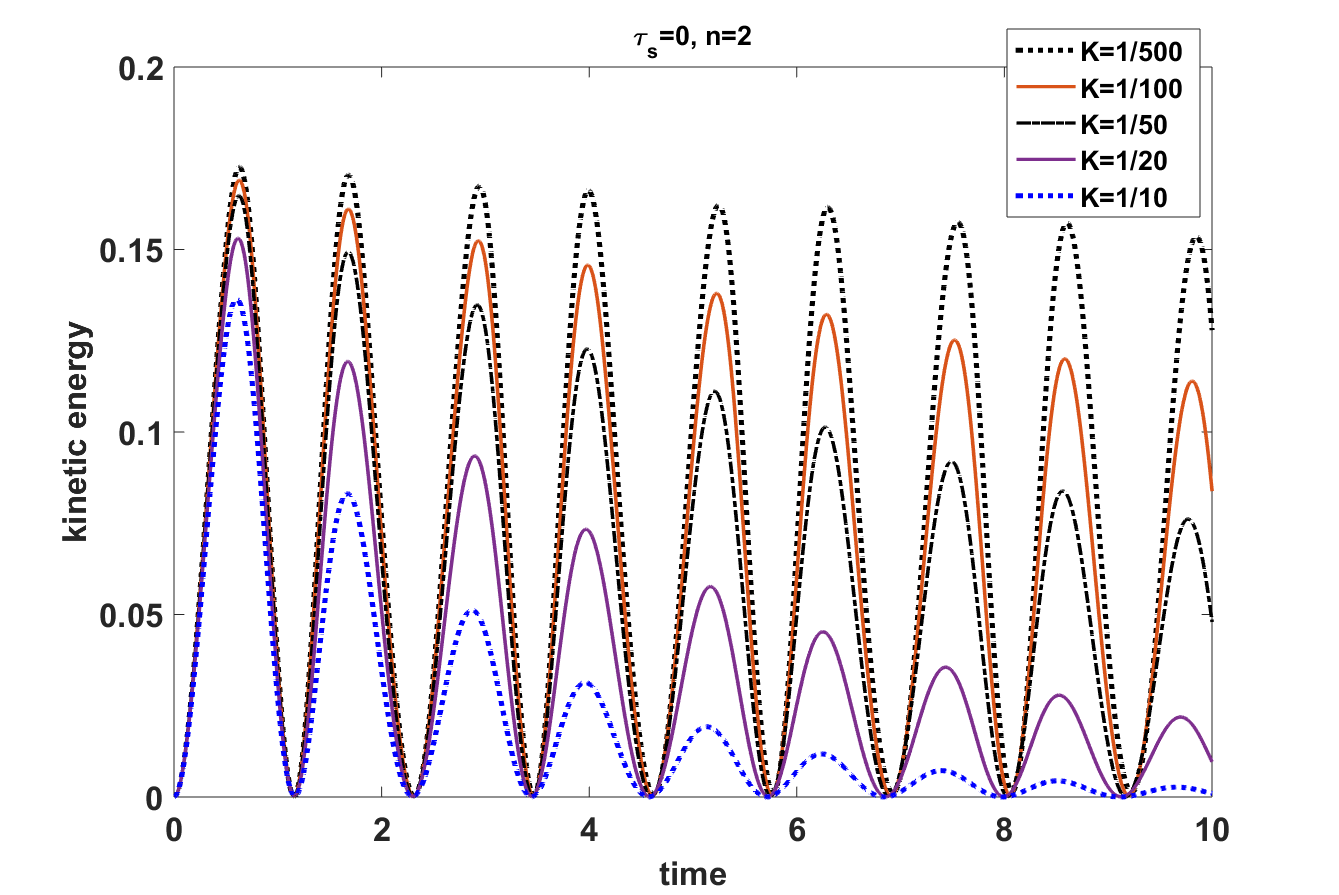}
 \includegraphics[width=.49\textwidth,natwidth=1369,natheight=905]{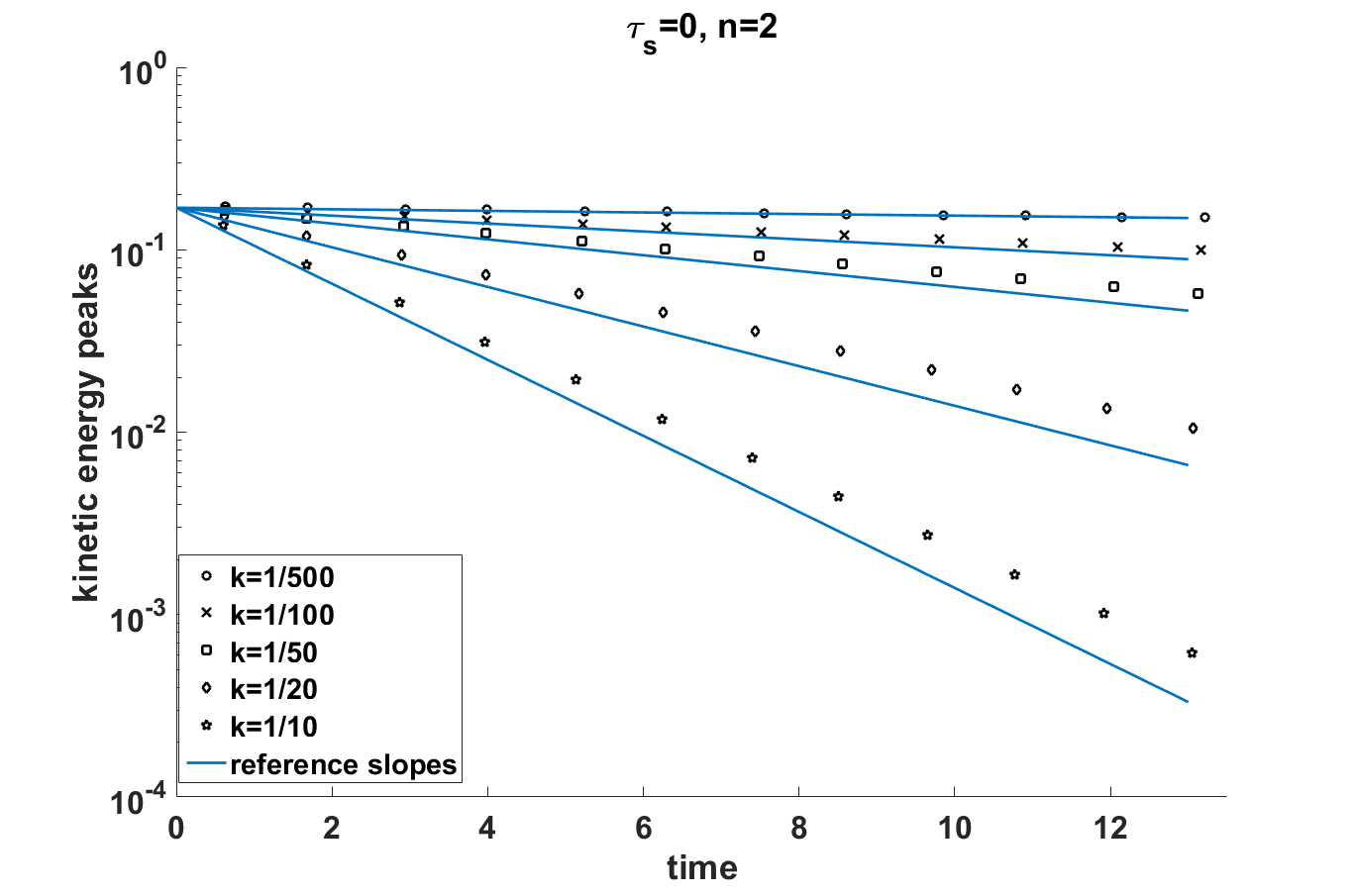}
 \caption{\label{fig0} Evolution of $E_{\rm kin}(t)$ and decay of the maximum (over a period) kinetic energy for the simulation of a Newtonian ($\tau_s=0,$ $\alpha=1$)  fluid droplet for several values of the viscosity parameter.}
\end{figure}

The computational domain in this and all further experiments is the cube $(0,\ell)^3$, $\ell=\frac{10}3$; an initially perturbed sphere of radius $r_0= 1$ is placed in the center of $\Omega$. Everywhere in computations we set $\rho=1$ and $\gamma=1$ (density and surface tension parameters), {while varying} $K$, $\tau_s$, $\alpha$, and $n$ (consistency, yield stress, flow index parameters and the number $n$ of spherical harmonic $\mathcal{H}_n$ used to set the initial perturbation). The perturbation parameter in \eqref{initial} is chosen to be $\tilde{\ep}=0.3$.
Further we use a sequence of descritizations with the following parameters: The maximal mesh size is $h_{\rm max}=\frac\ell{16}$; the mesh is aggressively refined towards the free surface, where the mesh size equals  $h_{\rm min}$.  Table~\ref{t0} shows the approximate number  of the degrees of freedom in the resulting descritizations (this number slightly varies when the free surface evolves) for different $h_{\rm min}$.

First we perform  a series of experiments for the \textit{Newtonian} oscillating droplet.
Although the numerical method was previously verified on a number of benchmark problems for Newtonian and viscoplastic fluids flows, the purpose of this experiment {is to assess the accuracy of the numerical method} and to study the convergence of flow statistics in this case to those given by the analysis {in \cite{Lamb,miller1968oscillations}} and recovered in \eqref{E_total} and \eqref{decay}. Thus, a droplet of the ideal fluid ($K=0$, $\tau_s=0$)  oscillates infinitely with constant amplitude. The deviation of the numerical solution from this behaviour allows us to estimate (by fitting an exponential function to maximum values of the  kinetic energy over periods) the numerical dissipation of the method, which is reported in Table~\ref{t0}. We see that the numerical dissipation is low and decreases when the mesh is refined.  All further experiments are done with $h_{\rm min}=\frac\ell{64} $. Figure~\ref{fig0} shows the evolution of the kinetic energy and the kinetic energy peaks for several values of the viscosity parameter. For reference, we plot the exponent functions from \eqref{decay} (there graphs are straight lines in the log scale). The {slopes} show the {theoretically predicted \textit{asymptotic} energy decay rates. Note that in the viscous case the  rate \eqref{decay} is valid for large enough time or sufficiently small perturbation $\tilde{\ep}$, see \cite{prosperetti1980free}, and so a deviation at the initial stage of oscillations may be expected.   The asymptotic rate}   is well predicted by the results of simulations. Therefore, we now turn to the numerical study of the yield stress case.

\begin{figure}[th]
 \centering
 \includegraphics[width=.49\textwidth,natwidth=1339,natheight=895]{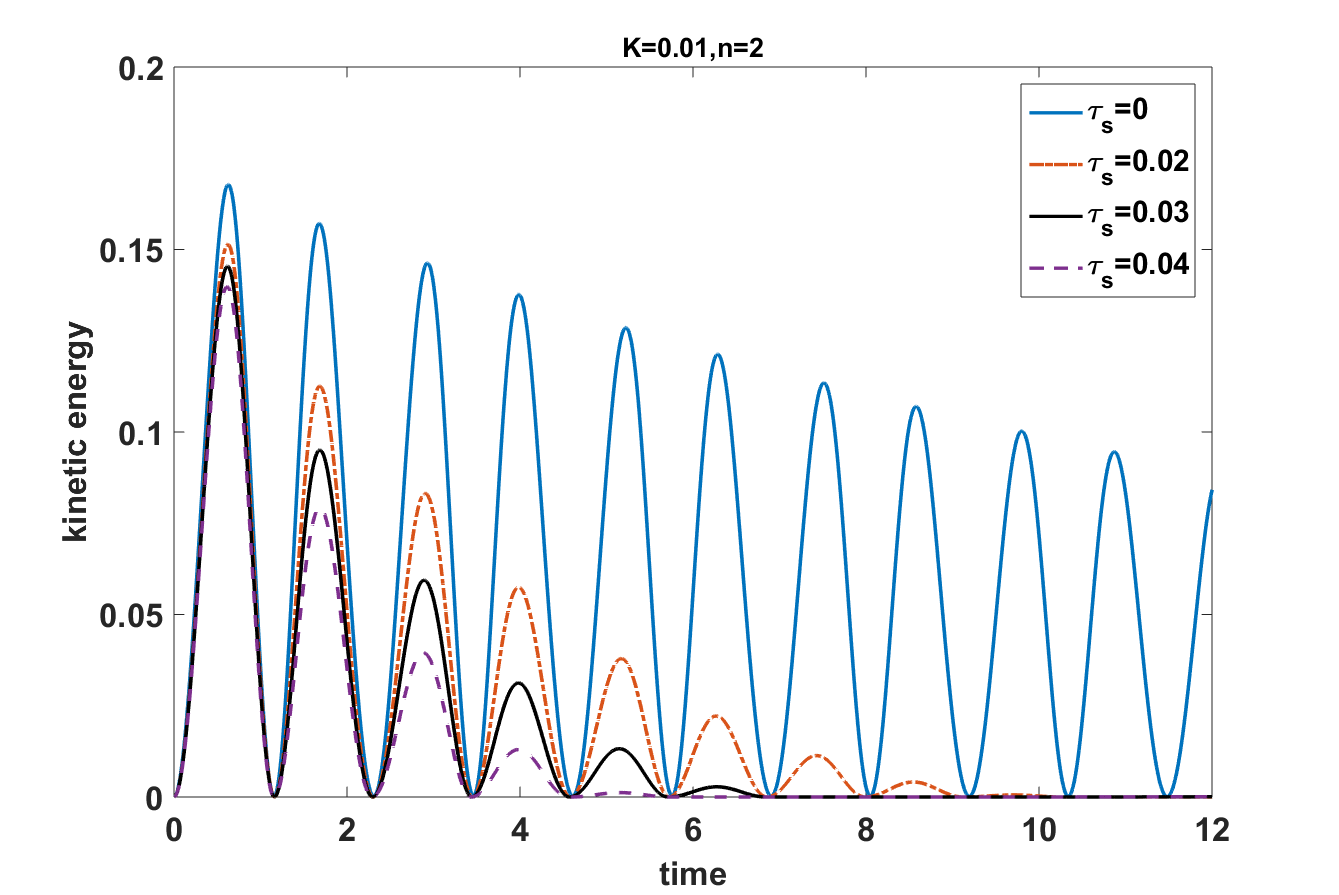}
 \includegraphics[width=.49\textwidth,natwidth=1339,natheight=895]{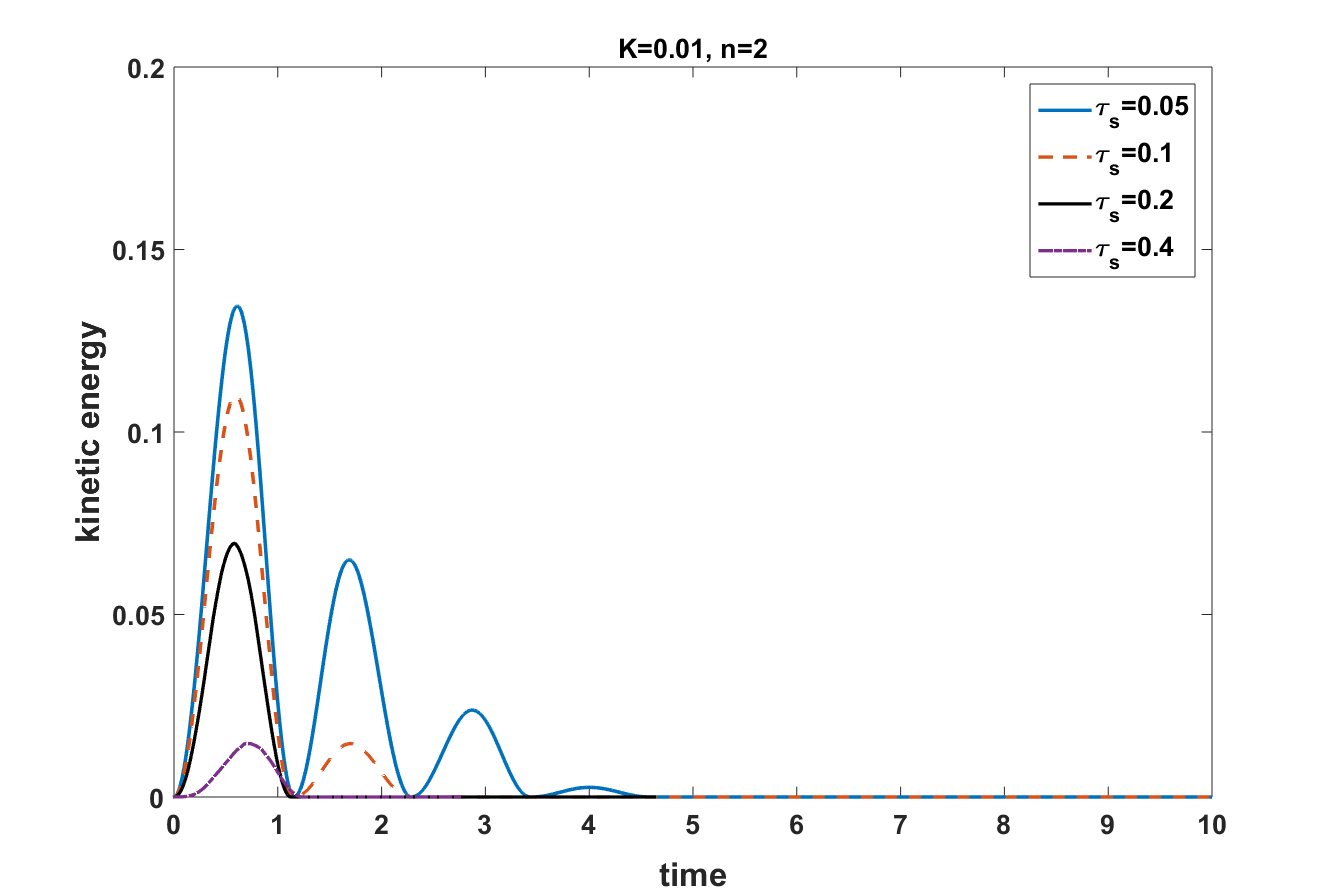}
 \includegraphics[width=.49\textwidth,natwidth=1339,natheight=895]{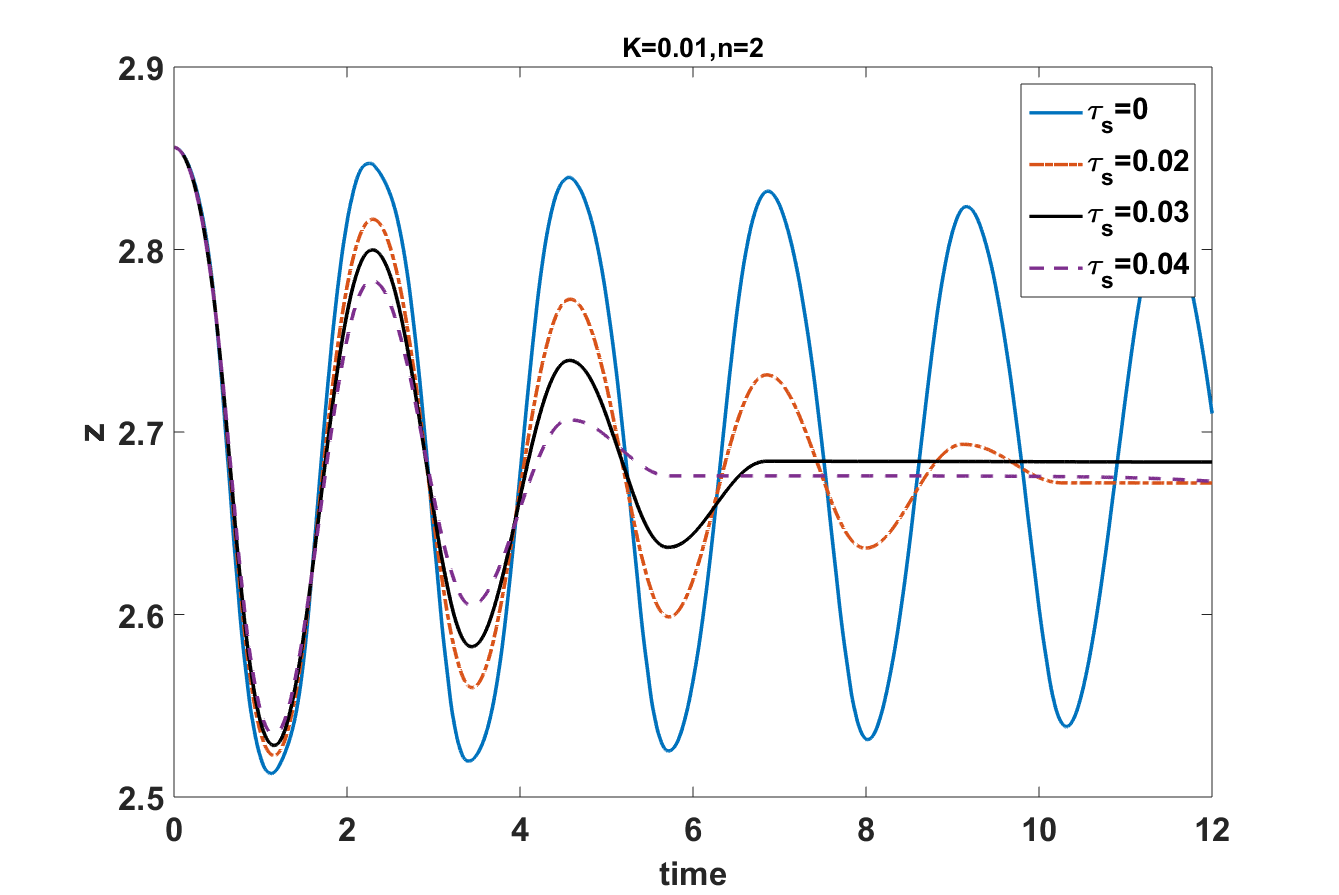}
 \includegraphics[width=.49\textwidth,natwidth=1339,natheight=895]{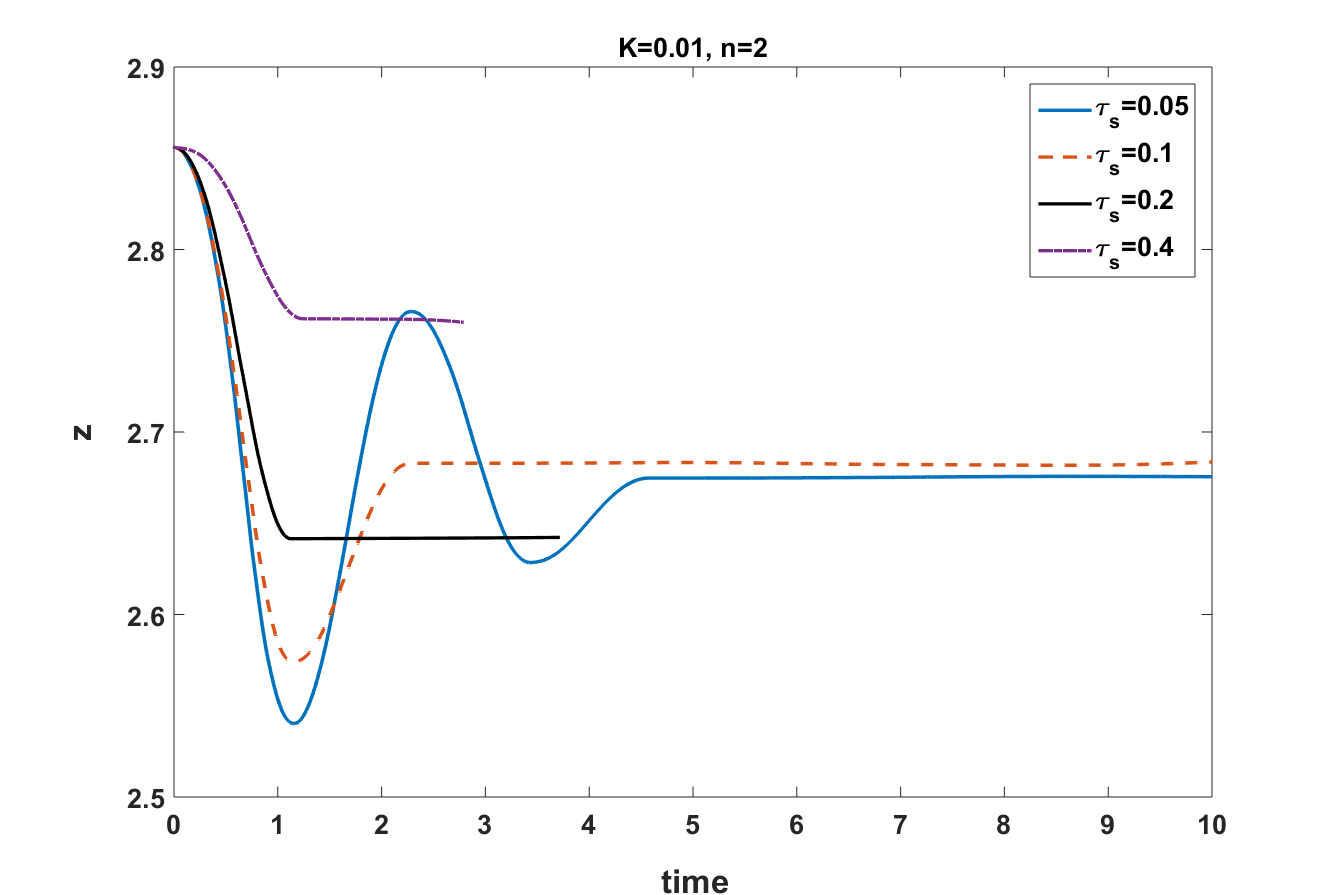}
 \caption{The evolution of the kinetic energy (upper plots) and the trajectory of the north tip (bottom plots) computed for $K=0.01$, different $\tau_s$ and the second spherical harmonic ($n=2$) to define initial perturbation.
 \label{fig1}}
\end{figure}

\begin{figure}[th]
 \centering
 \includegraphics[width=.49\textwidth,natwidth=1339,natheight=895]{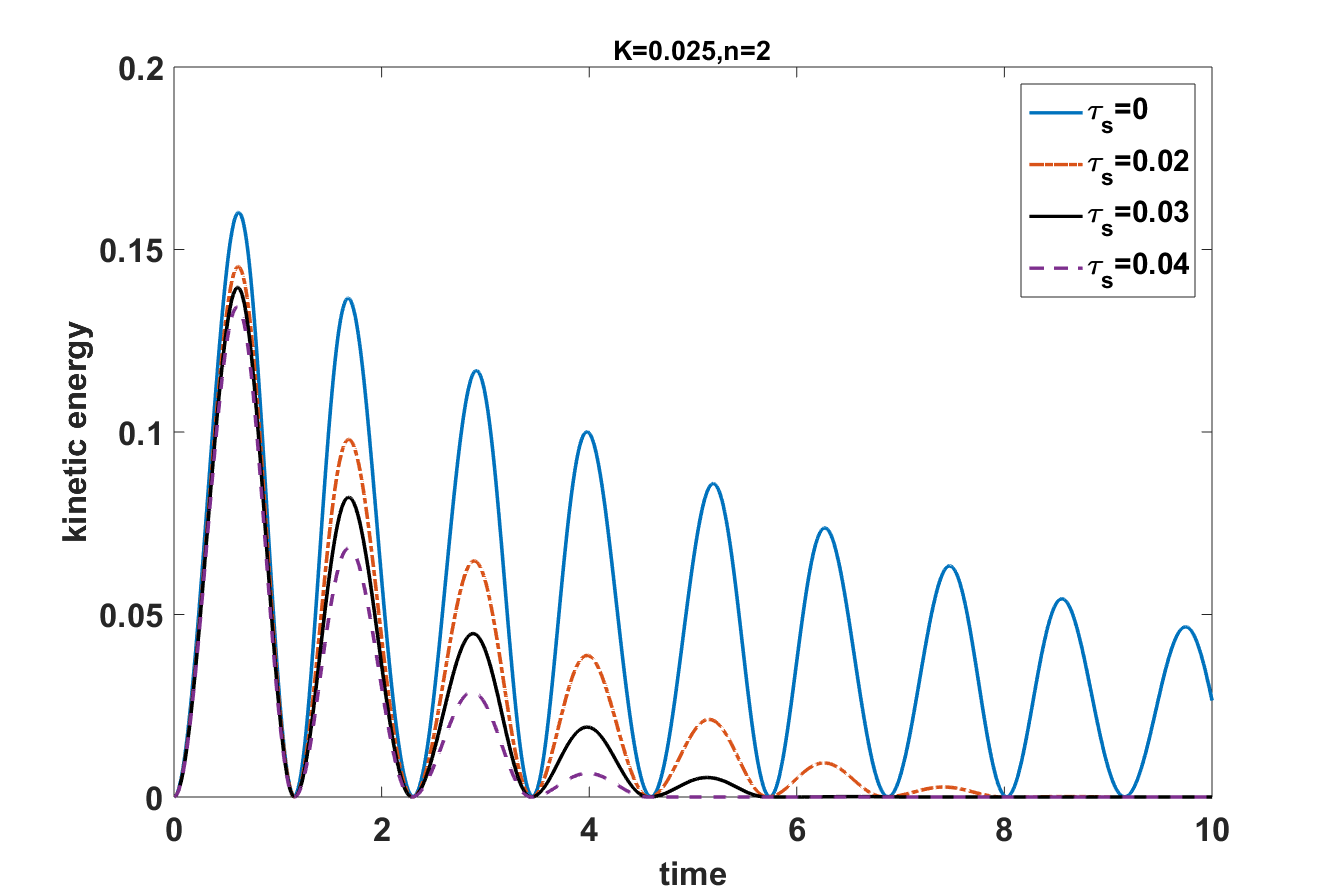}
 \includegraphics[width=.49\textwidth,natwidth=1339,natheight=895]{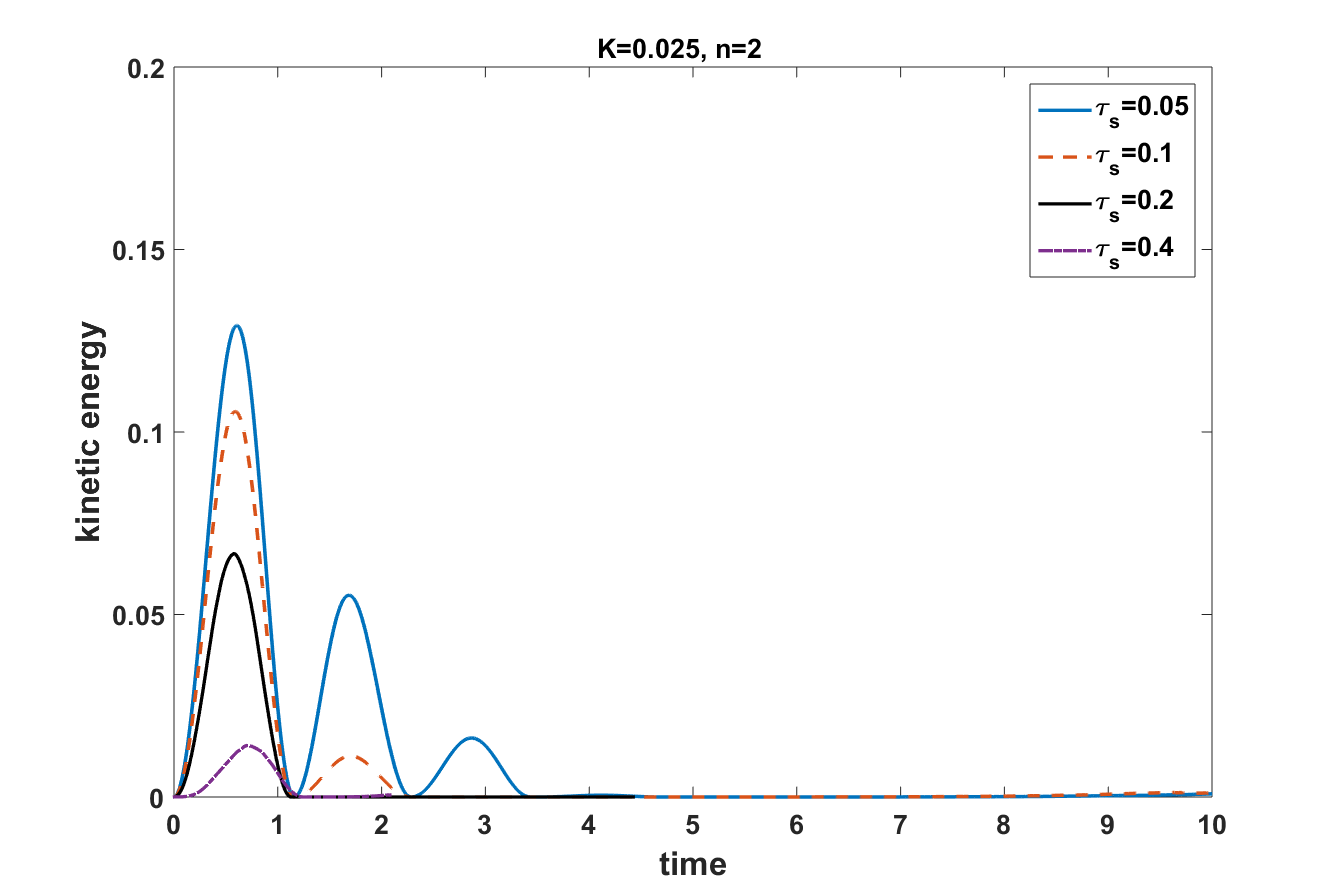}
 \includegraphics[width=.49\textwidth,natwidth=1339,natheight=895]{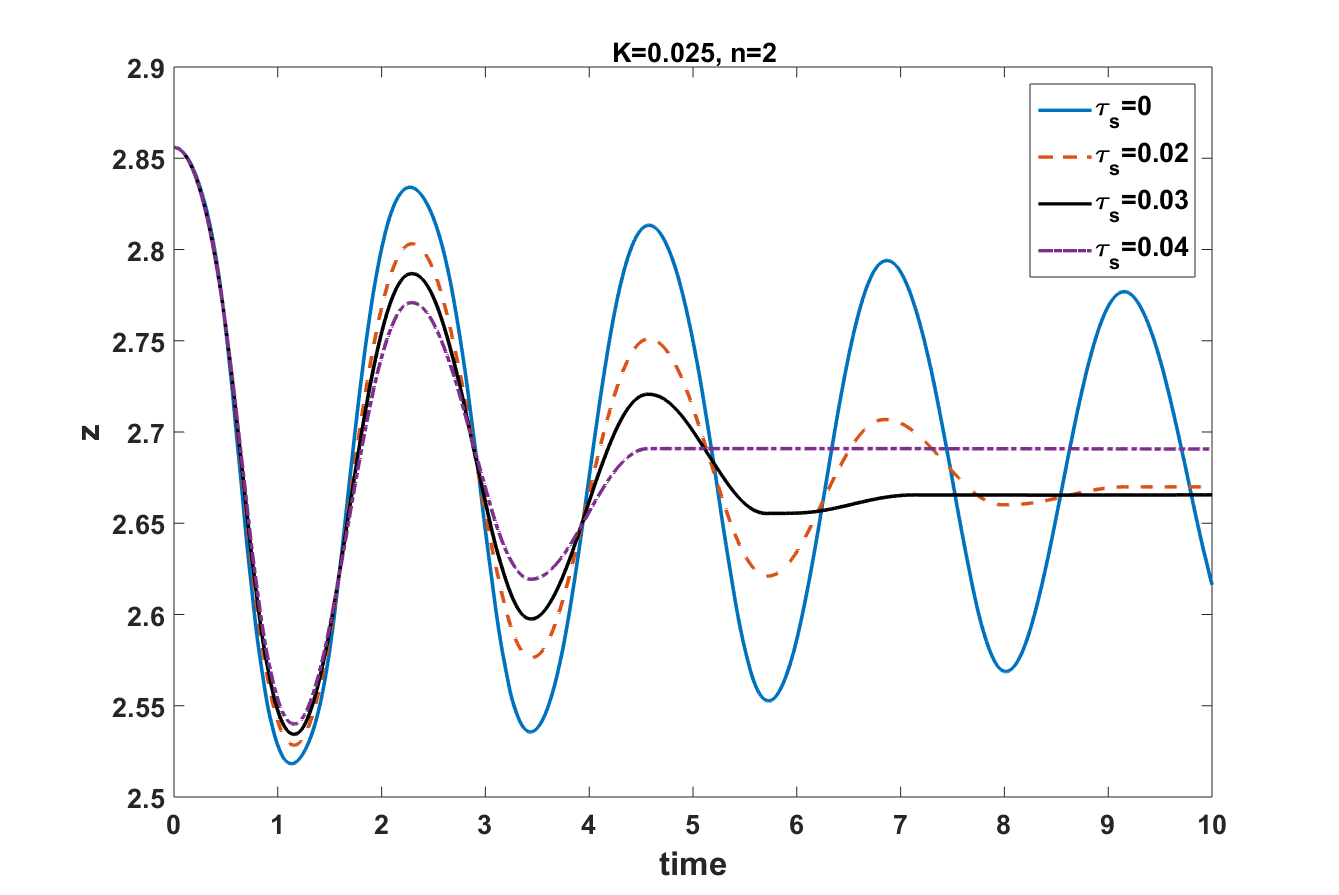}
 \includegraphics[width=.49\textwidth,natwidth=1339,natheight=895]{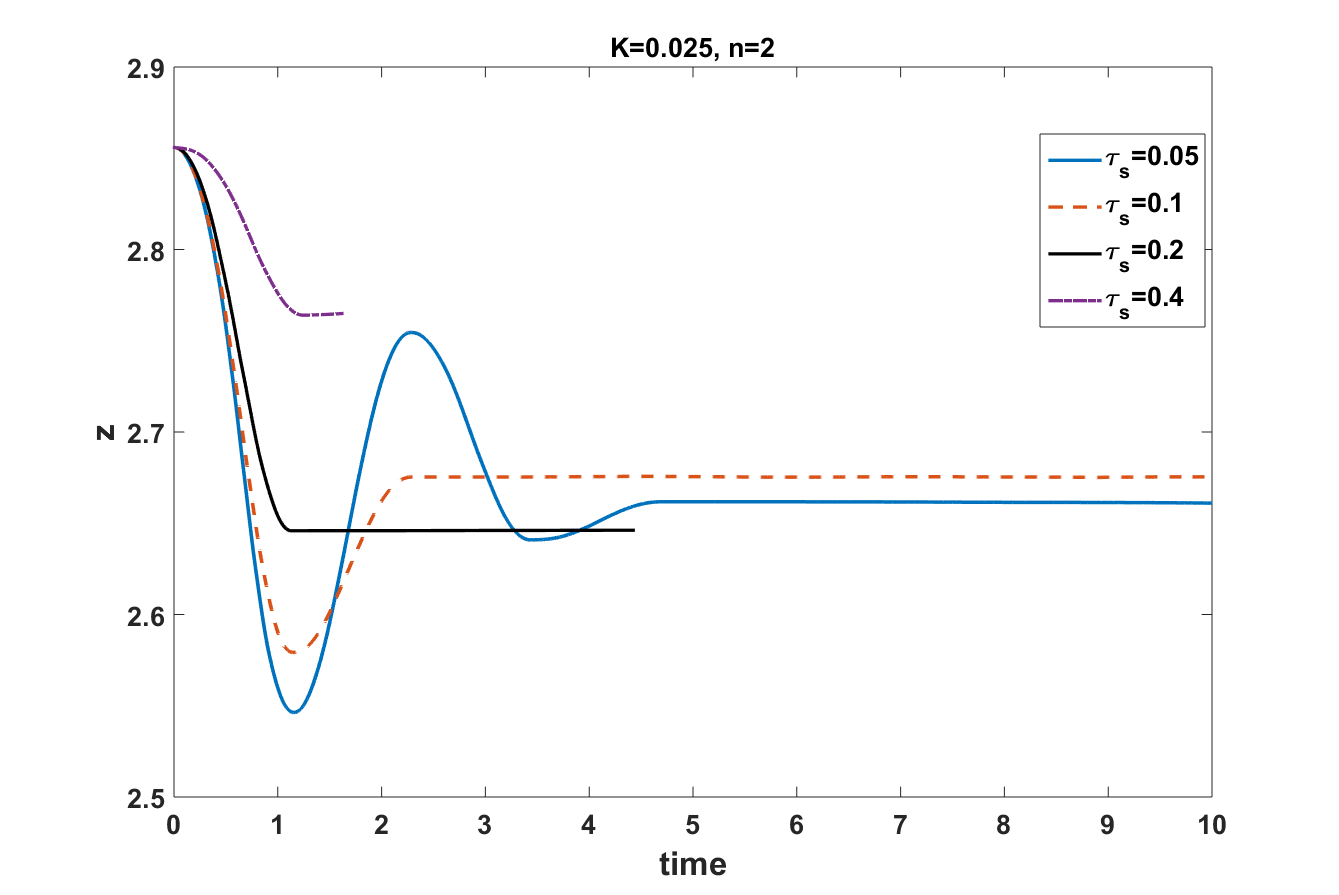}
 \caption{The evolution of the kinetic energy (upper plots) and the trajectory of the north tip (bottom plots) computed for $K=0.025$, different $\tau_s$ and the second spherical harmonic ($n=2$) to define initial perturbation.
 \label{fig2}}
\end{figure}

\begin{table}[th]
\centering
\caption{Computed and `predicted' final stopping times for various values of $K$ and $\tau_s$, with $\alpha=1$, $n=2$ (for initial perturbation).}
\label{t2}
\begin{tabular}{r|ll|ll|ll}
\hline
$\tau_s$&\multicolumn{2}{c}{K=0.005} & \multicolumn{2}{c}{K=0.01} & \multicolumn{2}{c}{K=0.025}\\
       &$T_f$& $T_{\rm pred}$  & $T_f$& $T_{\rm pred}$& $T_f$& $T_{\rm pred}$\\
\hline
 0.02    &   10.29  &  9.940    &  10.27   &  8.873  &  9.182 &   7.760  \\
0.03     &   8.274  &  6.647    &  6.847   &  6.629  &   7.125 &  5.543        \\
0.04     &   5.712  &  4.439    &  5.722   &  4.434  &  4.568  &  4.434  \\
0.05     &   4.573  &  4.416    &  4.577   &  4.361  &  4.688  &  3.326    \\
0.1       &   2.279  &  2.213    &  2.279   &  2.213  &   2.289 &  2.213 \\
0.2       &   1.135  &  1.099    &  1.130   &  1.099  &   1.130 &  1.099 \\
0.4       &   1.260  &   --          &  1.265  &  --         &   1.237        &  -- \\
\hline
\end{tabular}
\end{table}

First we experiment with the Bingham fluid (fluid index $\alpha=1$). As in the experiments with the Newtonian fluid, the  initial perturbation is
defined by \eqref{initial} with $A_2(0)=1$ and $A_n(0)=0$ for $n\neq2$, $\tilde{\ep}=0.3$. Now Figures~\ref{fig1}--\ref{fig2} show the evolution of the total kinetic energy and the trajectory of the north tip computed for viscosity coefficients $K=0.01$ and $K=0.025$ and different yield stress parameters $\tau_s$, with $\tau_s=0$ obviously showing the Newtonian case. Both from the kinetic energy evolution and the trajectory of the drop tip we clearly see the complete cessation of the motion in a finite time for all $\tau_s>0$. It is  interesting to note from the north tip trajectories that the final arrested state
is not necessarily the original unperturbed sphere. The quasi-period of the oscillations looks  independent of the $K$ and $\tau_s$ values. The decay rate and the final stopping time, otherwise, depend on  $K$ and $\tau_s$. The final stopping times  presented in Table~\ref{t2} were estimated from the computed kinetic
energy applying the following formula:
\begin{equation}\label{stop_Tf}
T_f=\text{arg}\min_{t>0}\max_{s\ge0}\{E_{\rm kin}(t+s)\le 10^{-7}\}.
\end{equation}
As can be expected $T_f$, in general, decreases for larger values of $K$ and $\tau_s$. It is interesting to note that
for the range of modest,  i.e., not too large, yield stress parameter values,  the final stoping time demonstrates
the dependence on $\tau_s$ close to $T_f=O(\tau_s^{-1})$.  The viscosity coefficient for this problem appears to have
less influence on the variation of the finite cessation time.
\smallskip

It follows from the analysis in section~\ref{s_drop}, and was noticed already in \cite{Lamb}, that for the Newtonian case the
drop oscillations are the linear superposition of individual oscillations of each spherical harmonics, satisfying equations \eqref{Anewton}.
For the {non-Newtonian} case, we do not see why a similar superposition principle should  be valid in general.
However, if for a prediction purpose one could \textit{assume} that there is no transfer of energy between different scales, then
one can write an ODE for the time evolution of each harmonic separately. In addition to the terms in \eqref{Anewton} one computes for the
plastic dissipation: $\tau_s\int_{\Omega(t)}|\mathbf{D}\bu|\dx\simeq \tau_s n^{-1}r_0^{1-n}\left|\frac{dA_n}{dt}\right|\int_{\Omega_0}|D^2(r^n \mathcal{H}_n)|\dx$, where $D^2(f)$ is the Hessian matrix for $f$.  For example, for the second spherical harmonic the corresponding
ODE reads
\begin{equation}\label{ODE2}
\frac{d^2A_2}{dt^2}+\frac{ 5K}{r_0^{2}\rho}\frac{dA_2}{dt}+ \frac{ 8\tau}{r_0^{3}\rho}A_2 + \frac{\tau_s}{r_0 \rho} \frac{10}{\sqrt{6}}\text{sign}\left(\frac{d A_2}{dt}\right)=0.
\end{equation}
{This} ODE can be numerically solved with a high accuracy. The first time $T_{\rm pred}$ such that $A_n(t)=0$ for all $t>T_{\rm pred}$
may serve {as} a  prediction to the actual stopping time if  the  initial perturbation is defined only by the second harmonic (similar with other harmonics).
We solve \eqref{ODE2} by the 4th order Runge-Kutta method and report the computed  $T_{\rm pred}$ in Table~\ref{t2}. The obtained   $T_{\rm pred}$
are close to $T_f$ recovered by full 3D simulations. This suggests that the transfer of energy between the scales (from lower to higher)
does not play an essential role in this problem and gives an additional support to the assumption leading to \eqref{aux2}. The ``---'' sign in Table~\ref{t2} indicates that for $\tau_s=0.4$ according to the ODE the drop motion is halted at $t=0$.
\smallskip

\begin{figure}[tp]
 \centering
 \includegraphics[width=.49\textwidth,natwidth=1339,natheight=895]{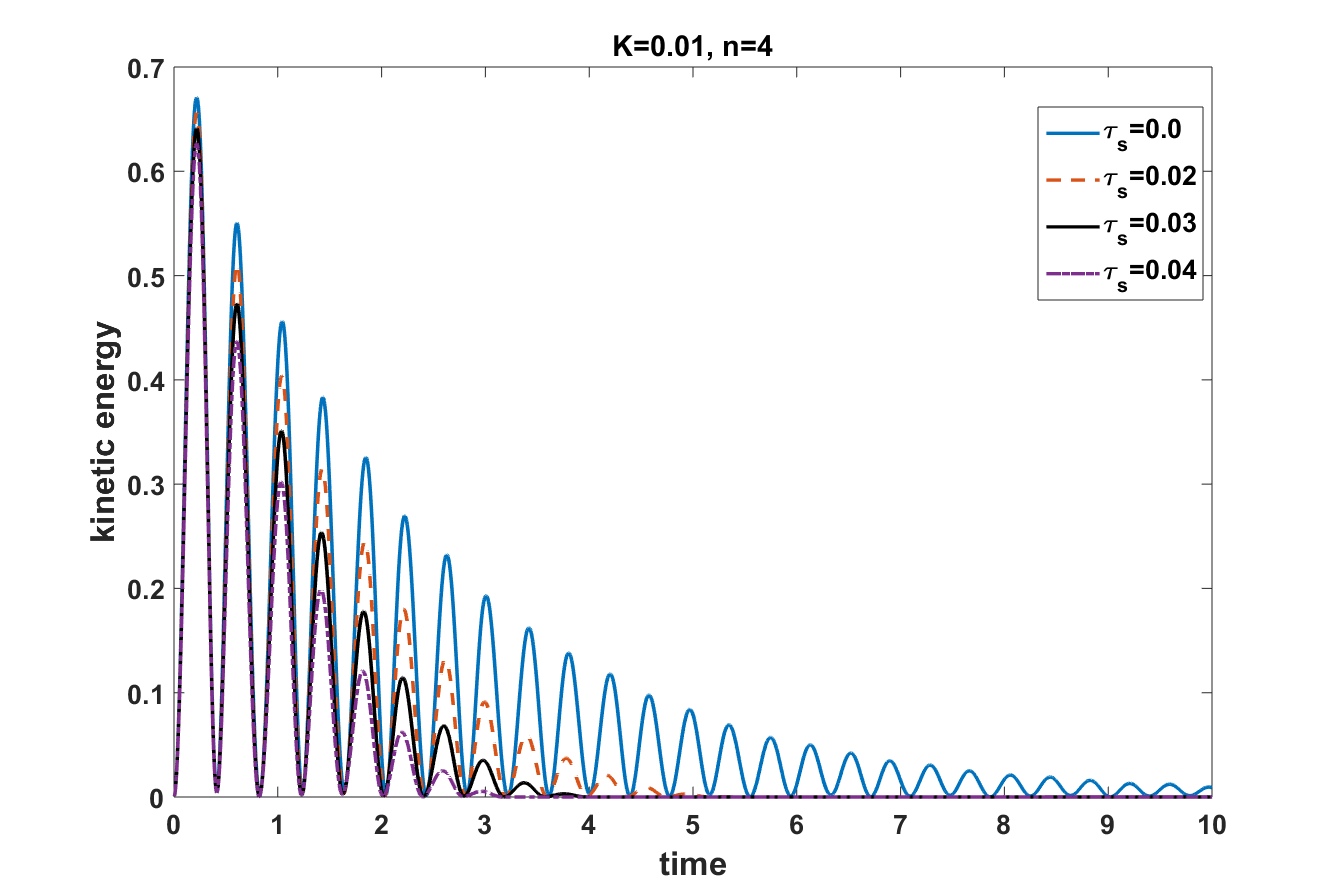}
 \includegraphics[width=.49\textwidth,natwidth=1339,natheight=895]{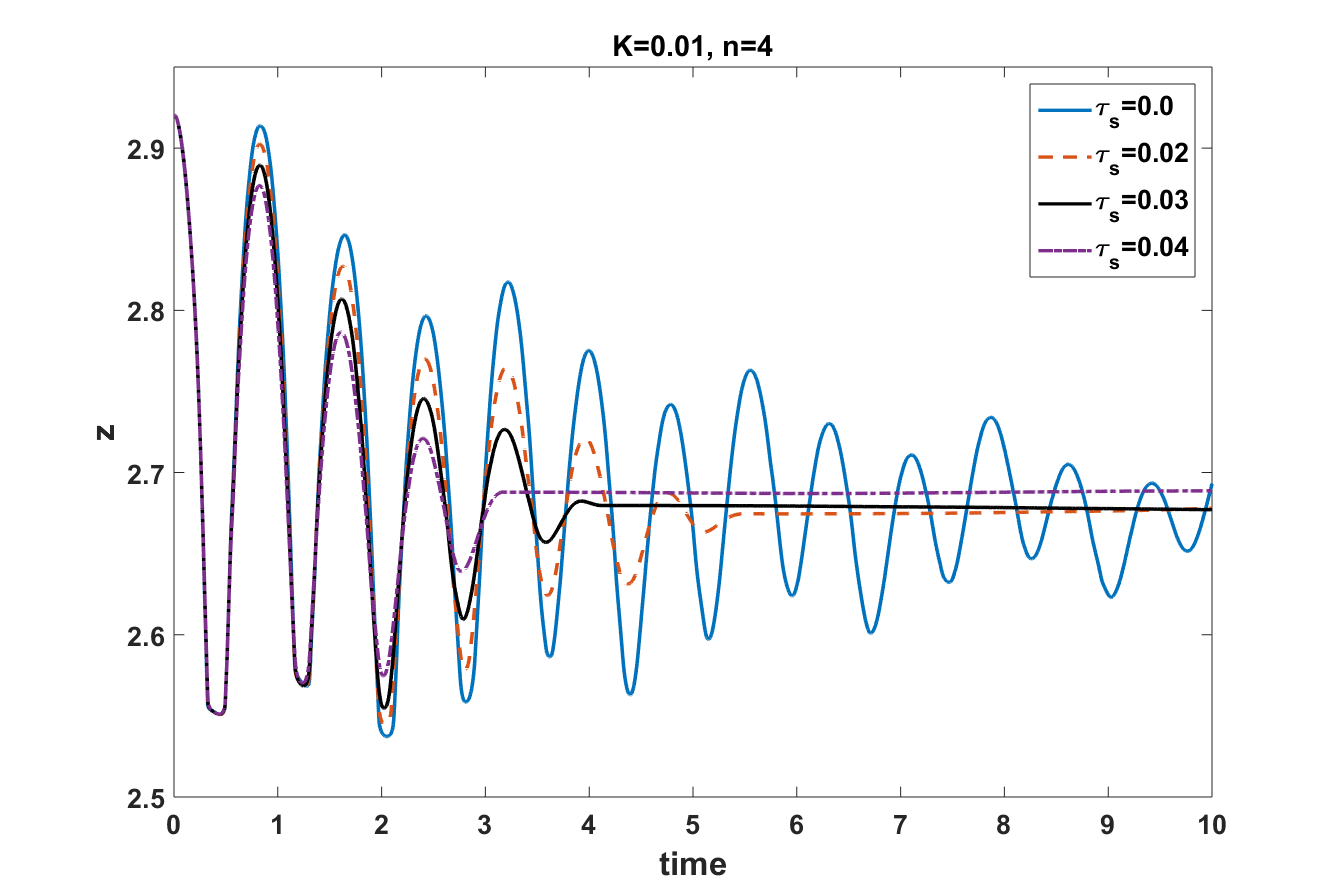}
 \caption{The evolution of the kinetic energy (left) and the trajectory of the north tip (right) computed for $K=0.01$ and various
 $\tau_s$, with $n=4$ (for initial perturbation).
 \label{fig4}}
\end{figure}

We next experiment with different initial perturbations of the drop. In this experiment, we set
$A_4(0)=1$ and $A_n(0)=0$ for $n\neq4$, $\tilde{\ep}=0.3$  in \eqref{initial}. For this  setup, Figure~\ref{fig4}
shows the evolution of the total kinetic energy and the trajectory of the north tip computed for viscosity coefficient $K=0.01$ and
 different yield stress parameters $\tau_s$. Again  we observe the complete cessation of the motion in a finite time for all $\tau_s>0$.
As  expected from the analysis, the decay of the oscillations for the spherical harmonic with larger number happens faster and the
computed stopping time $T_f$ is smaller.

\begin{figure}[tp]
 \centering
 \includegraphics[width=.49\textwidth,natwidth=1339,natheight=895]{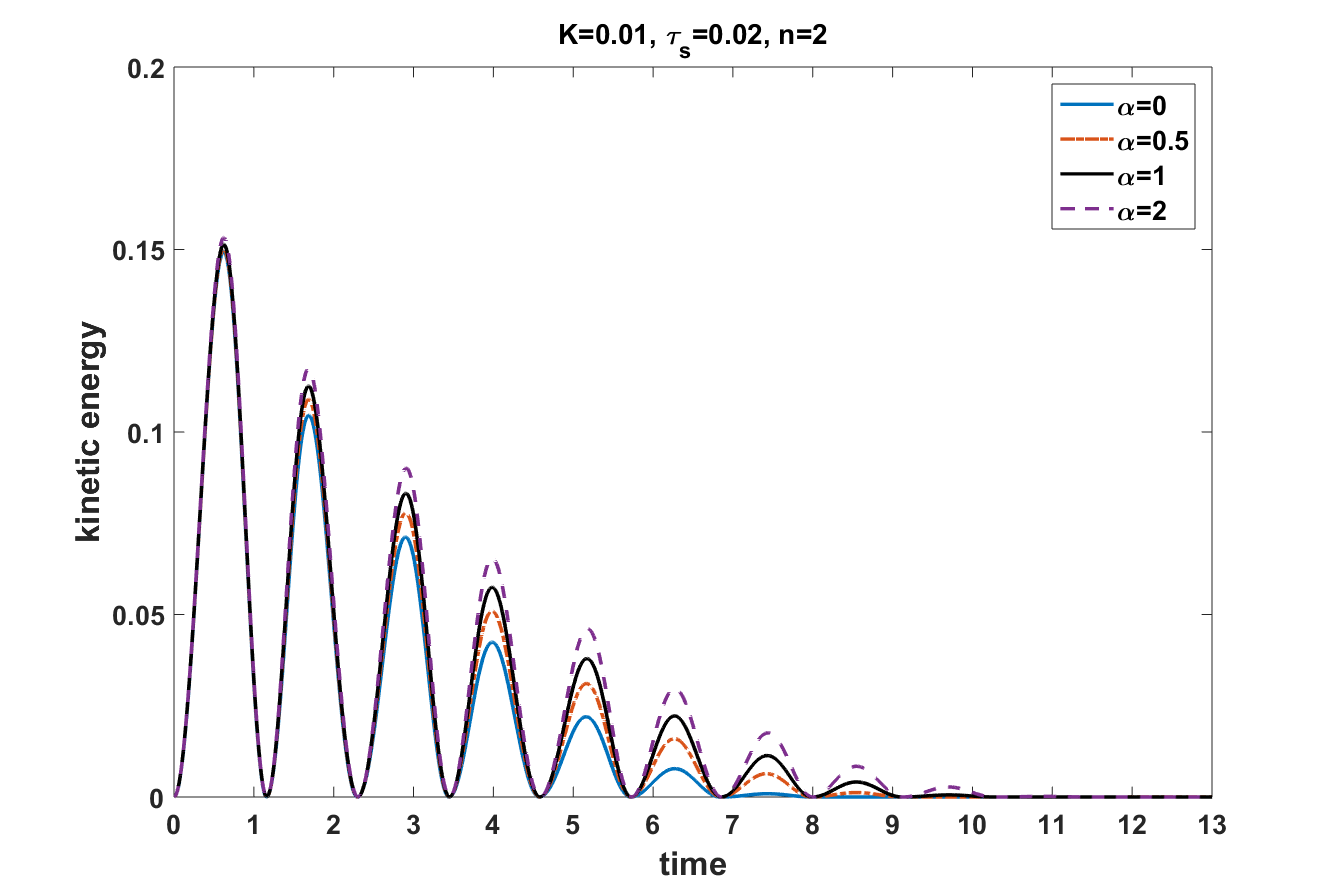}
 \includegraphics[width=.49\textwidth,natwidth=1339,natheight=895]{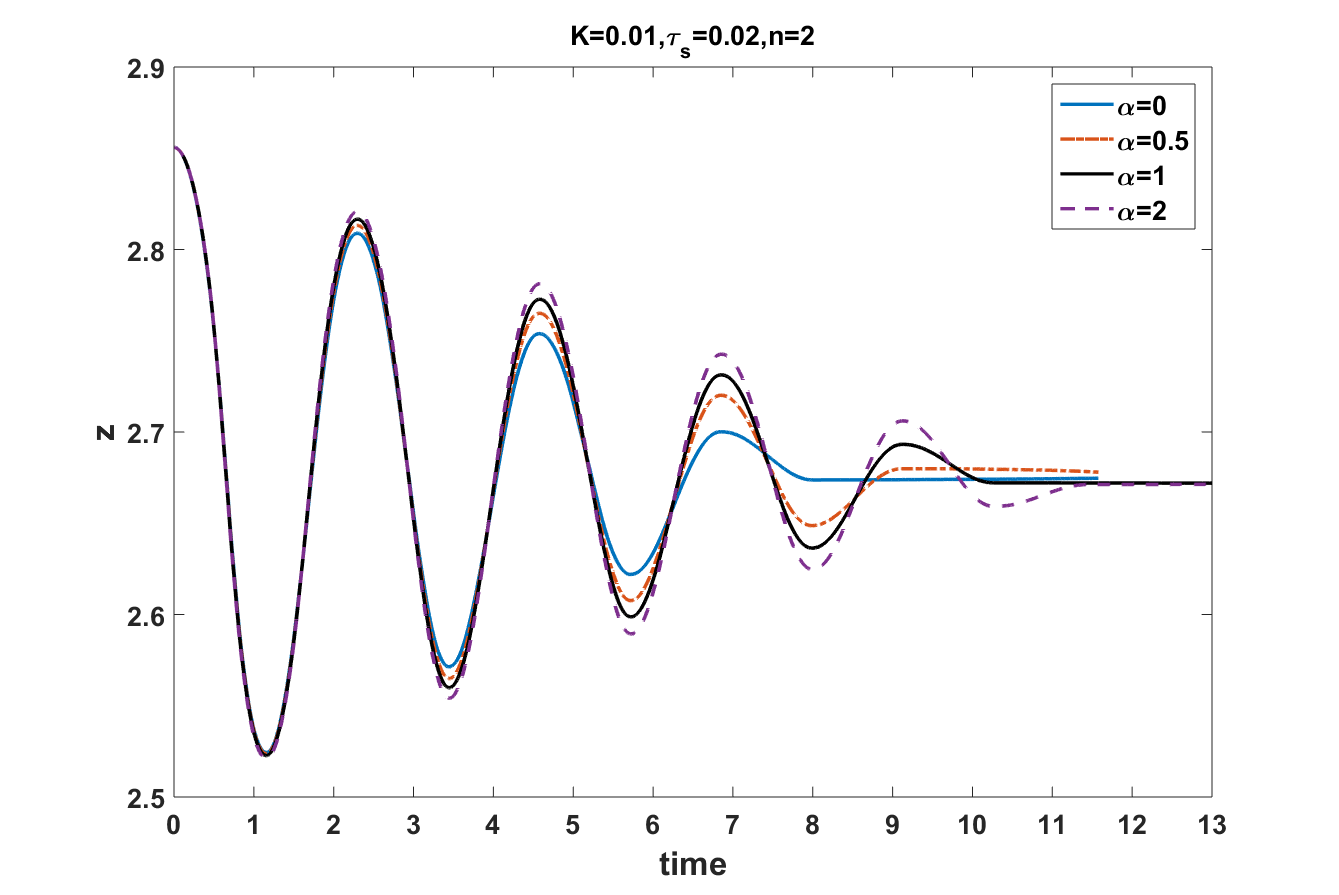}
 \caption{The evolution of the kinetic energy (left) and the trajectory of the north tip (right) computed for $K=0.01$,
 $\tau_s=0.02$, and different flow indexes $\alpha$.
 \label{fig3}}
\end{figure}

Further, we simulate the droplet oscillations for different values of the fluid index $\alpha$. The computed evolution of the total kinetic energy
and the trajectory of the north tip for  $K=0.01$ and $\tau_s=0.02$ are shown in Figure~\ref{fig3}. The estimated final stopping times are shown in
Table~\ref{t3}. The results indicate that shear-thinning/{thickening} variation  has some affect on the stopping times of the oscillations in general leading to
faster decay {as} $\alpha\to0$. At the same time, the results for $n=4$ are inconclusive.

\begin{table}[h]
\centering
\caption{Estimated final stopping times for various values of $\alpha$ and $n\in\{2,4\}$
(for initial perturbation), with $K=0.01$, $\tau_s=0.02$. \label{t3}}
\begin{tabular}{lclclclclc|c|}
\hline
$\alpha$&n=2 & n=4  \\
\hline
0    & 7.96&5.55&\\
0.5 &9.09&5.53&\\
1    &10.27&5.46&\\
2    &11.40&4.92&\\
\hline
\end{tabular}
\end{table}

\begin{figure}[th]
 \centering
 \includegraphics[width=.25\textwidth,natwidth=1136,natheight=840]{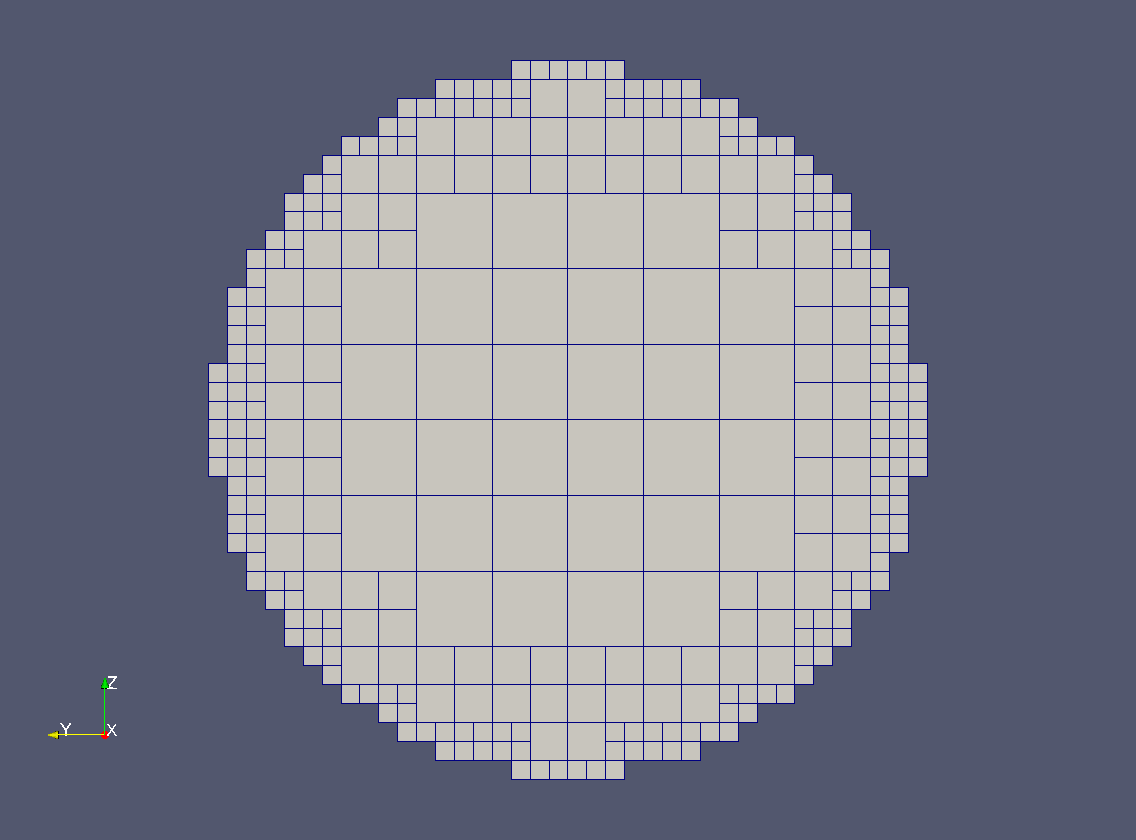}\qquad
 \includegraphics[width=.25\textwidth,natwidth=1136,natheight=840]{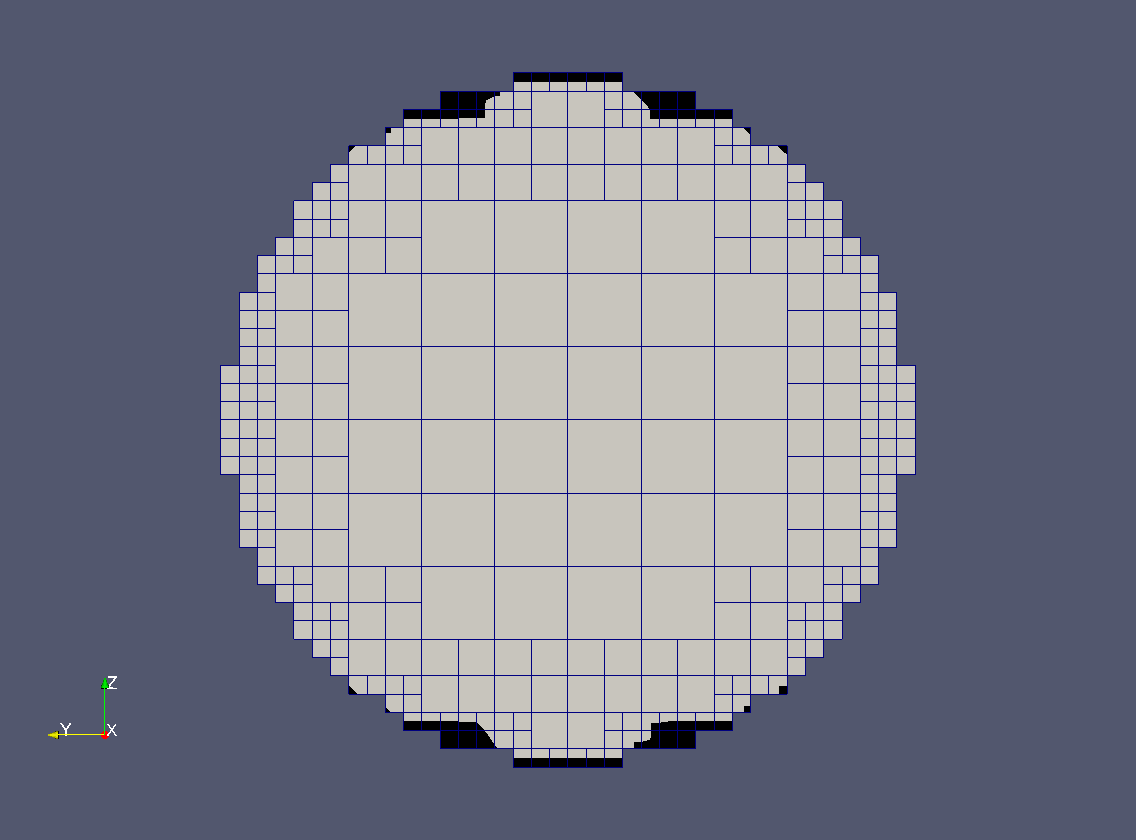}\qquad
 \includegraphics[width=.25\textwidth,natwidth=1136,natheight=840]{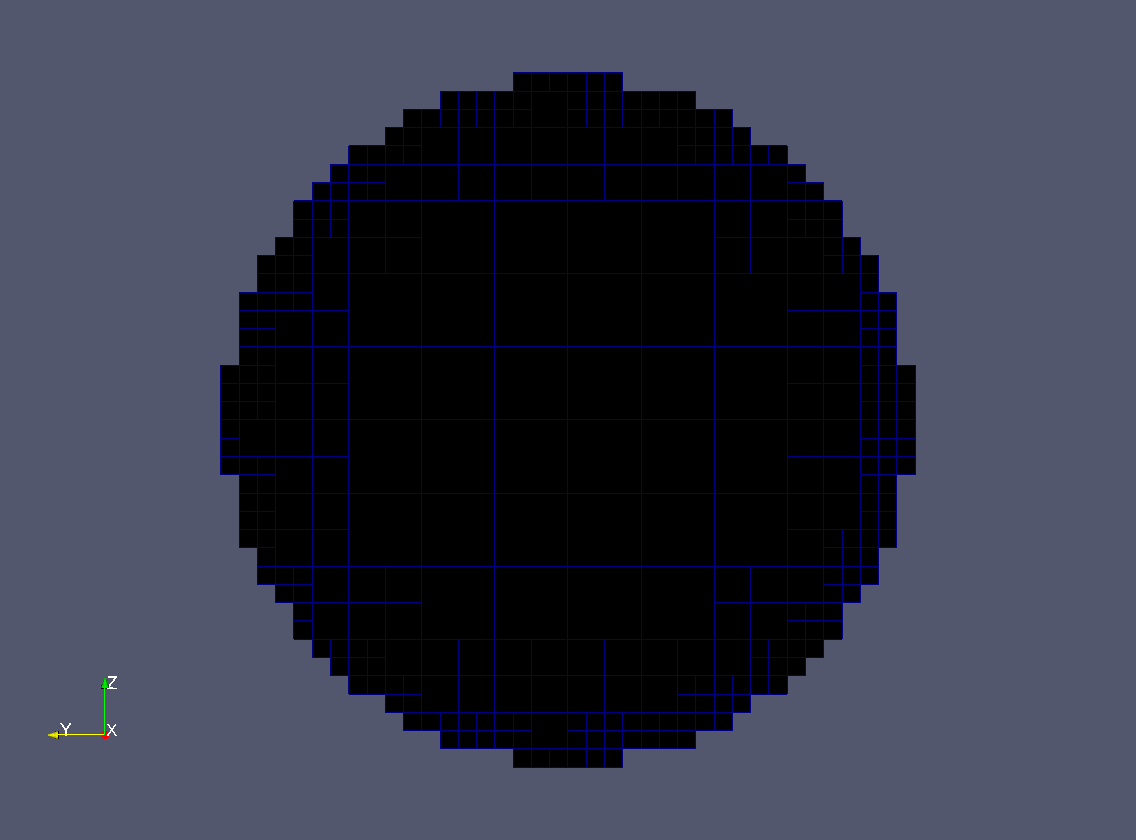}\\
 \begin{minipage}[1ex]{.25\textwidth}
 t=5.703
 \end{minipage}\qquad
 \begin{minipage}[1ex]{.25\textwidth}
 t=5.722
 \end{minipage}\qquad
 \begin{minipage}[1ex]{.25\textwidth}
 t=5.740
 \end{minipage}\\[2ex]
 \includegraphics[width=.25\textwidth,natwidth=1136,natheight=840]{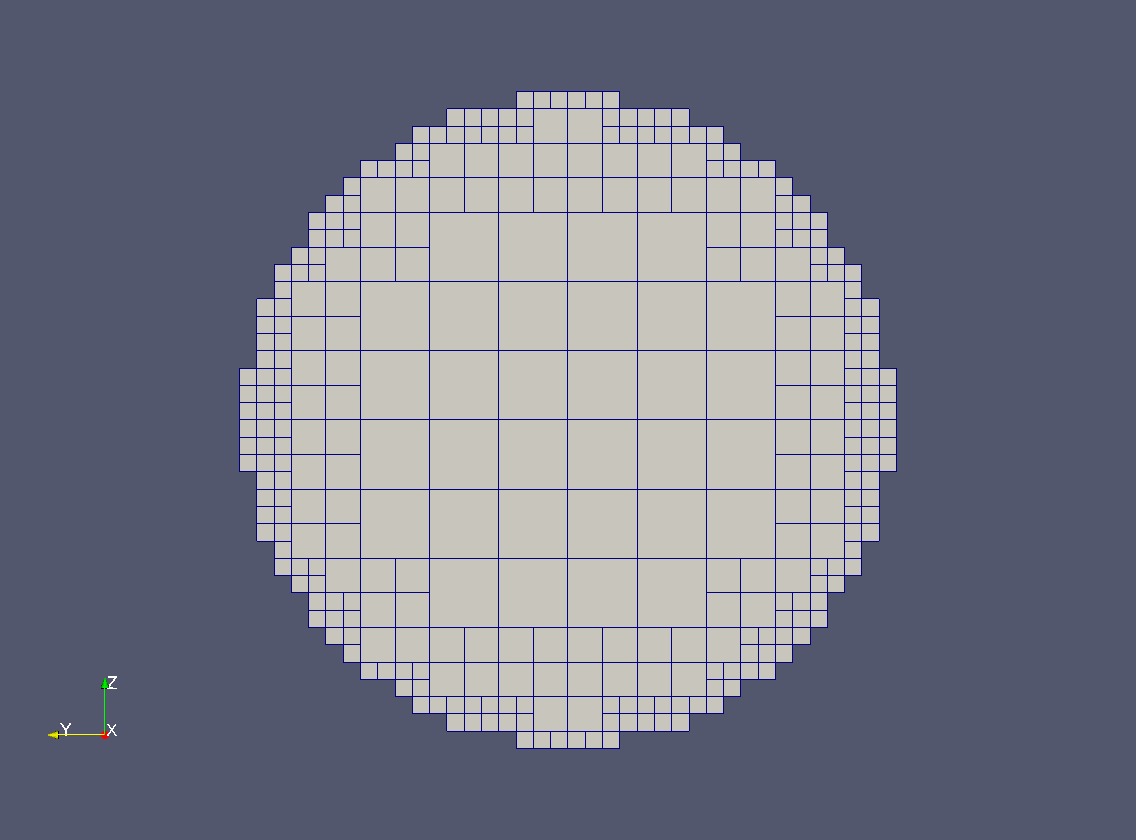}\qquad
 \includegraphics[width=.25\textwidth,natwidth=1136,natheight=840]{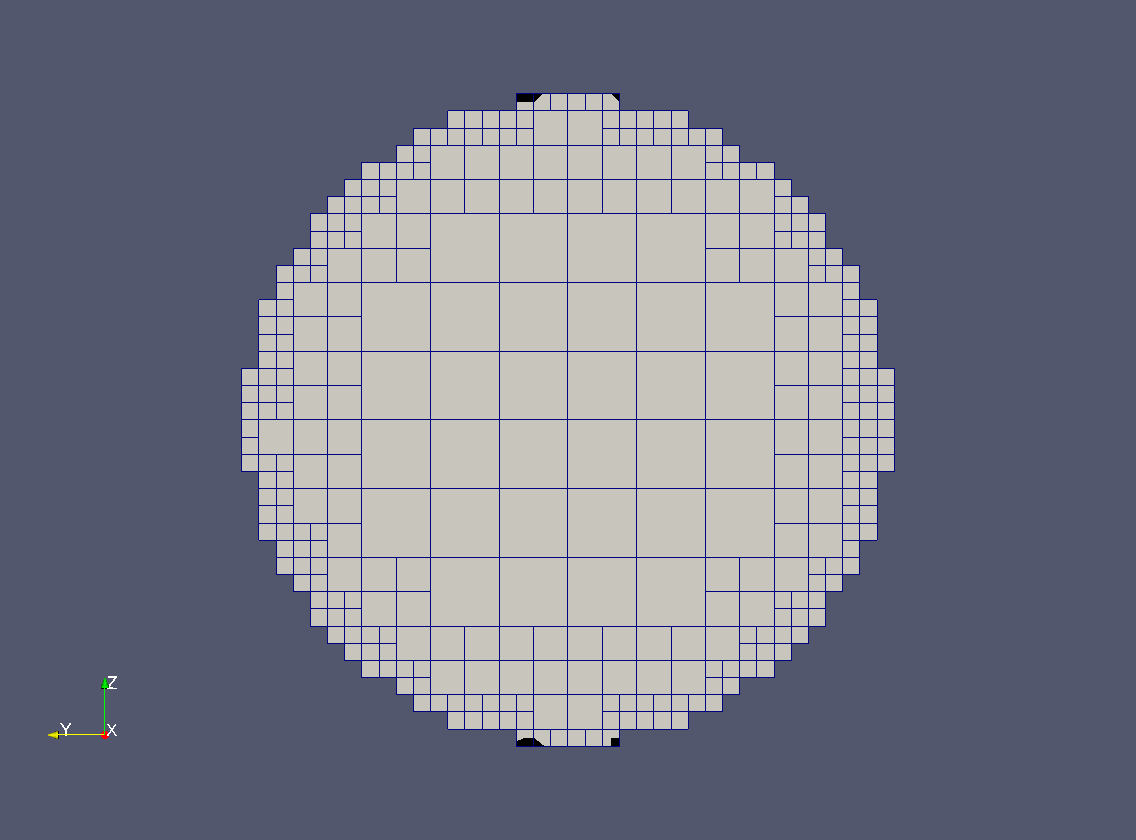}\qquad
 \includegraphics[width=.25\textwidth,natwidth=1136,natheight=840]{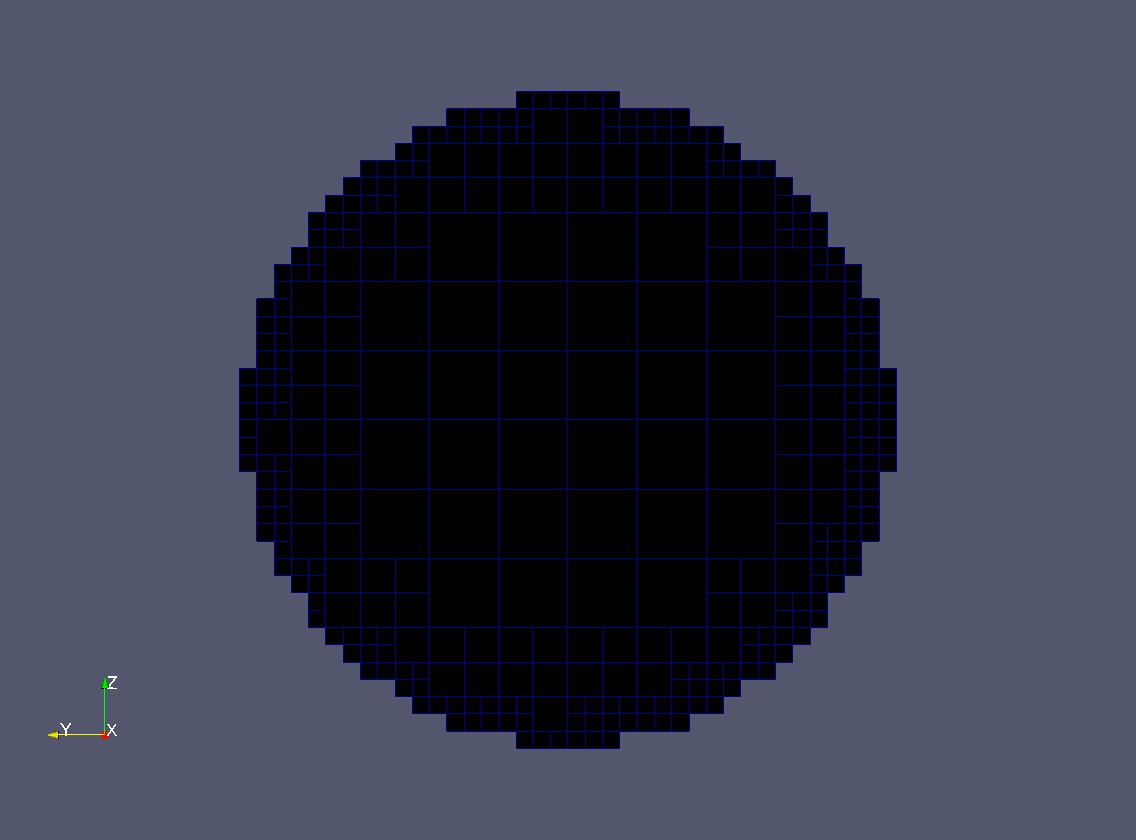}\\
 \begin{minipage}[1ex]{.25\textwidth}
 t=4.563
 \end{minipage}\qquad
 \begin{minipage}[1ex]{.25\textwidth}
 t=4.582
 \end{minipage}\qquad
 \begin{minipage}[1ex]{.25\textwidth}
 t=4.600
 \end{minipage}
 \caption{The visualization of the rigid zones near the final stopping time for $\tau_s=0.04$ (upper plots) and $\tau_s=0.05$ (bottom plots), with other parameters  $K=0.01$, $\alpha=1$, $n=2$. The cutaway by the xz-midplane is shown. The unyielded regions by von Mises criterion are colored black. Full cells are shown, but cut cells ensuring $O(h^2)$ boundary approximation are used in computations.  \label{fig_yeild}}
\end{figure}

We are also interested in the evolution of unyielded zones prior to the final cessation of the drop motion.
Note that numerical studies of the pipe and enclosed flows typically demonstrate an earlier formation and further growth
of the unyielded zones until they occupy the whole domain and halt the motion, see, e.g.,         \cite{chatzimina2005cessation,chatzimina2007cessation,syrakos2016cessation}. However,  for the oscillating drop problem,
if we accept the approach of Lamb and seek the solution in the form of the series \eqref{osc}, then we conclude that the whole droplet comes to the full stop at $T_f$ without \textit{prior} formation of rigid zones. The solution in \eqref{osc} is an approximation, and it is interesting to see which scenario the fluid motion follows in practice. Results of the numerical experiments suggest that Lamb's approach is remarkably predictive. Figure~\ref{fig_yeild} shows the  unyielded regions computed with the help of von Mises criterion
around the final stopping time for the Bingham fluid and with other parameters $K=0.01$, $\tau\in\{0.04,\,0.05\}$. The regions are visualized at three consecutive time steps. We see the (almost) immediate transition from fluidic to rigid phases in the entire droplet. Small unyielded regions near the droplet tips right before the complete stop can be a numerical
phenomenon. We recall that the numerical method makes \textit{no} use of the expansion in \eqref{osc} or any other
assumptions, including rotational symmetry, made in the framework of section~\ref{s_drop}; { rather, it obtains the 3D solution of  \eqref{N-S}--\eqref{b2} directly.} Postprocessing of the numerical results for other values of $\tau_s$ showed very similar behavior of the rigid zones to those shown in Figure~\ref{fig_yeild}, so we skip including these plots.
It also occurs that the von Mises criterion yields the final stopping times  very close to those computed from \eqref{stop_Tf}.

{
\begin{figure}[tp]
 \centering
 \includegraphics[width=.49\textwidth,natwidth=1091,natheight=730]{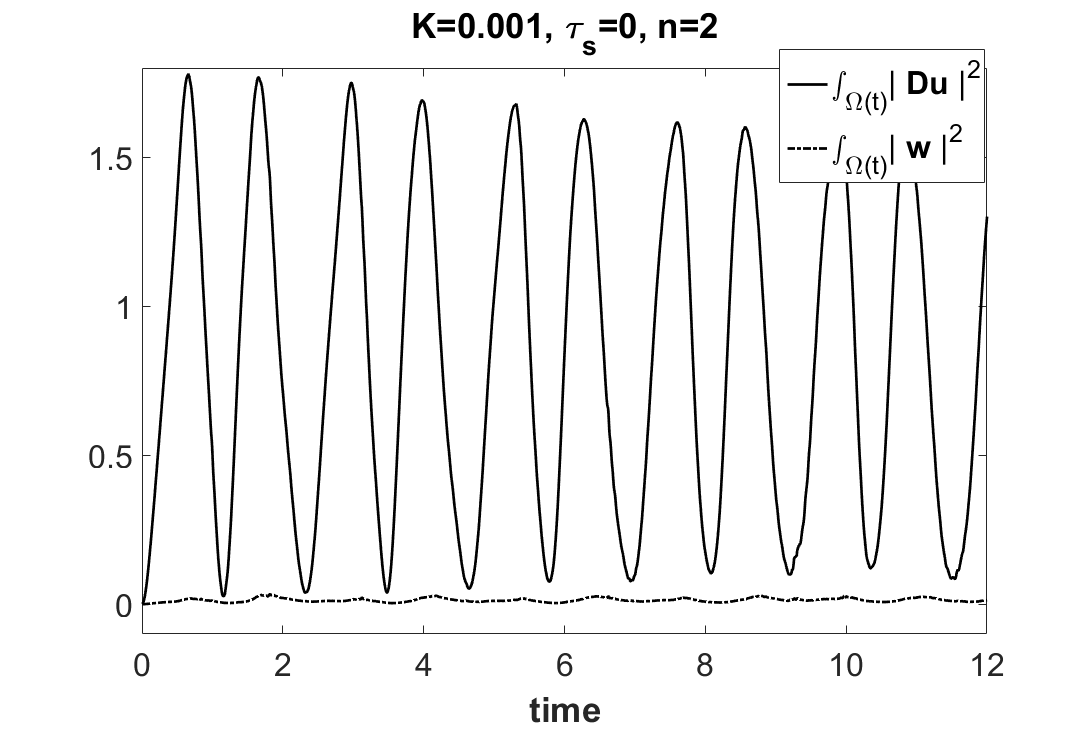}
 \includegraphics[width=.49\textwidth,natwidth=1369,natheight=905]{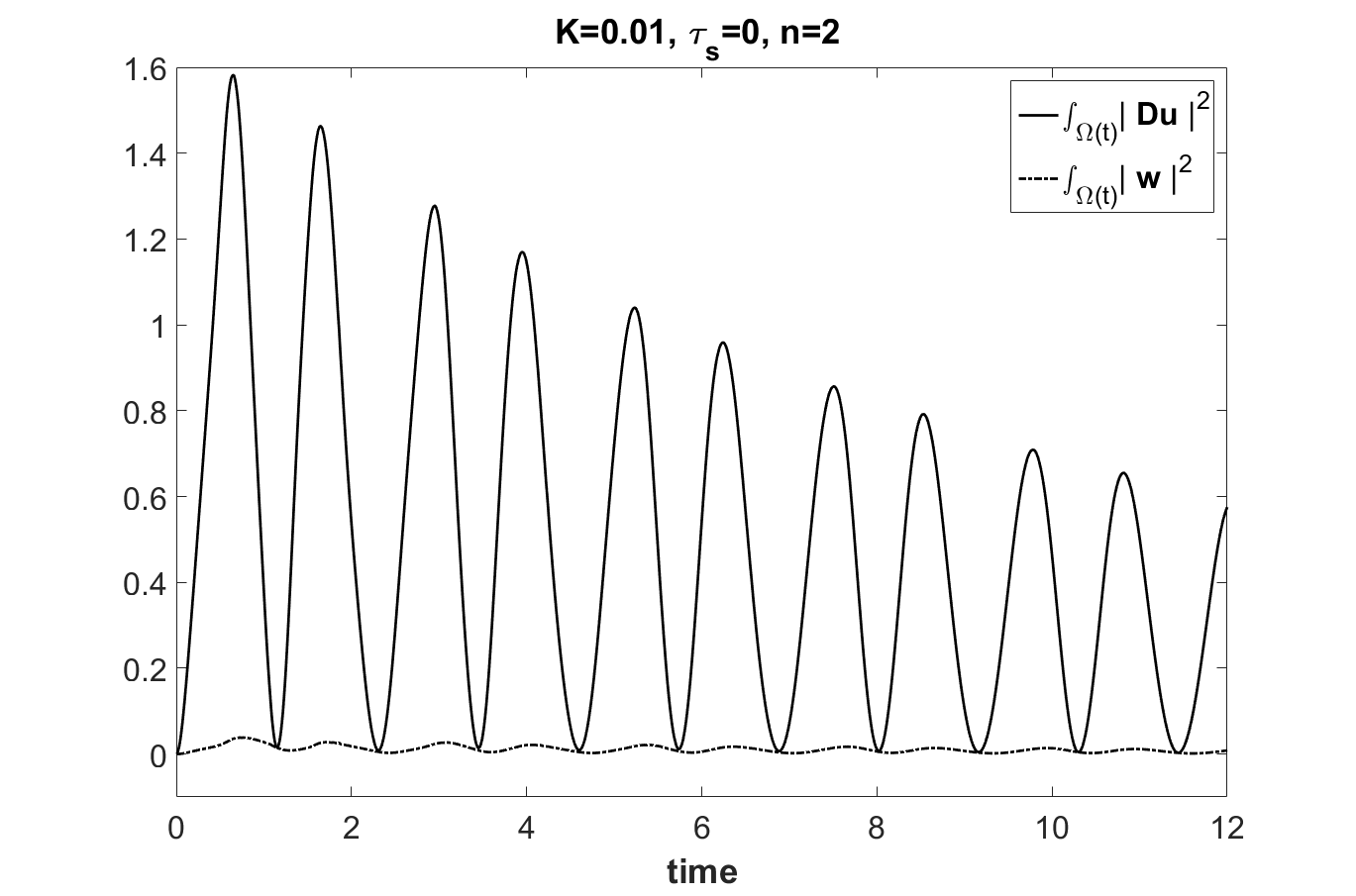}
 \includegraphics[width=.49\textwidth,natwidth=1369,natheight=905]{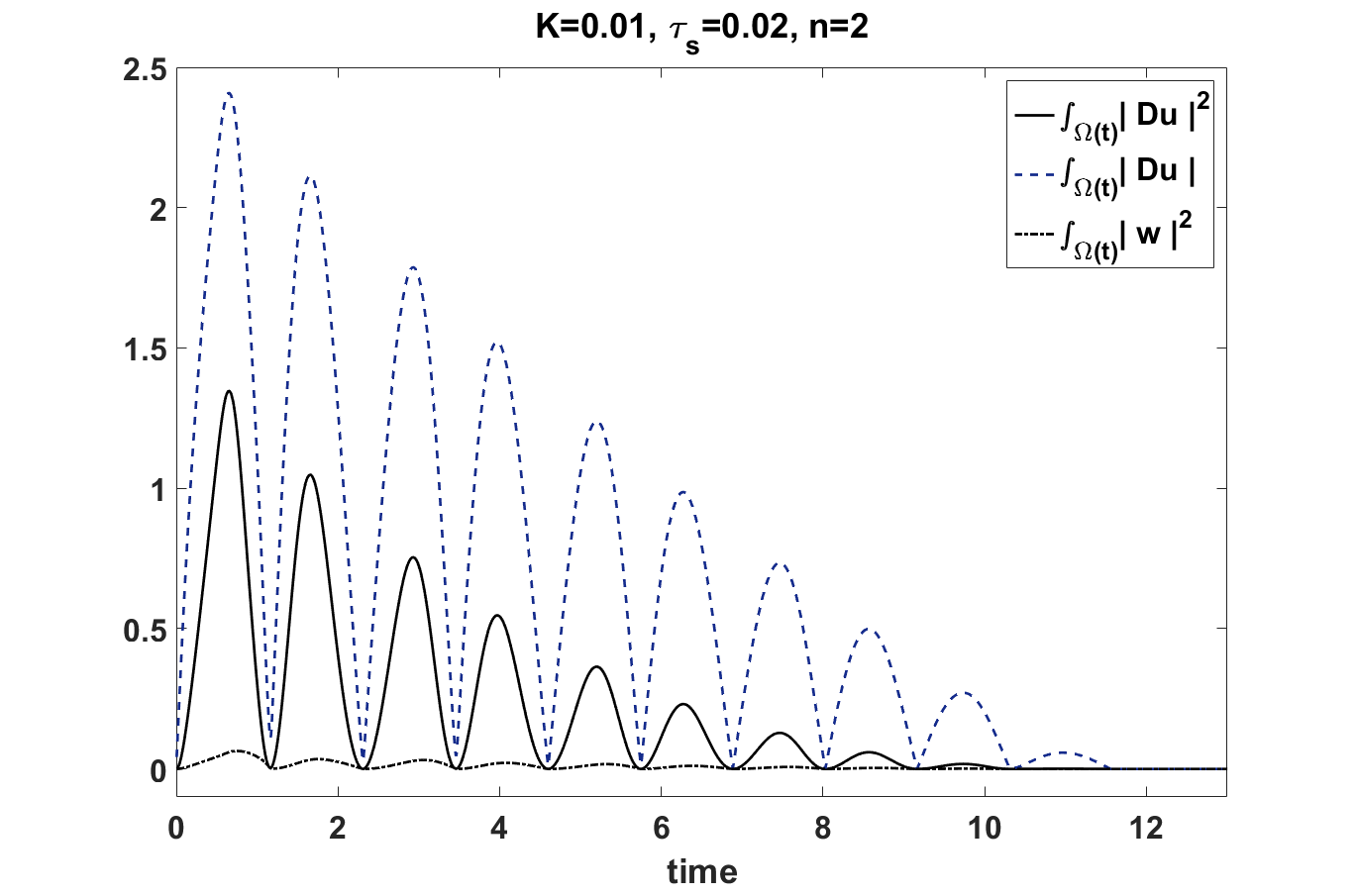}
 \includegraphics[width=.49\textwidth,natwidth=1369,natheight=905]{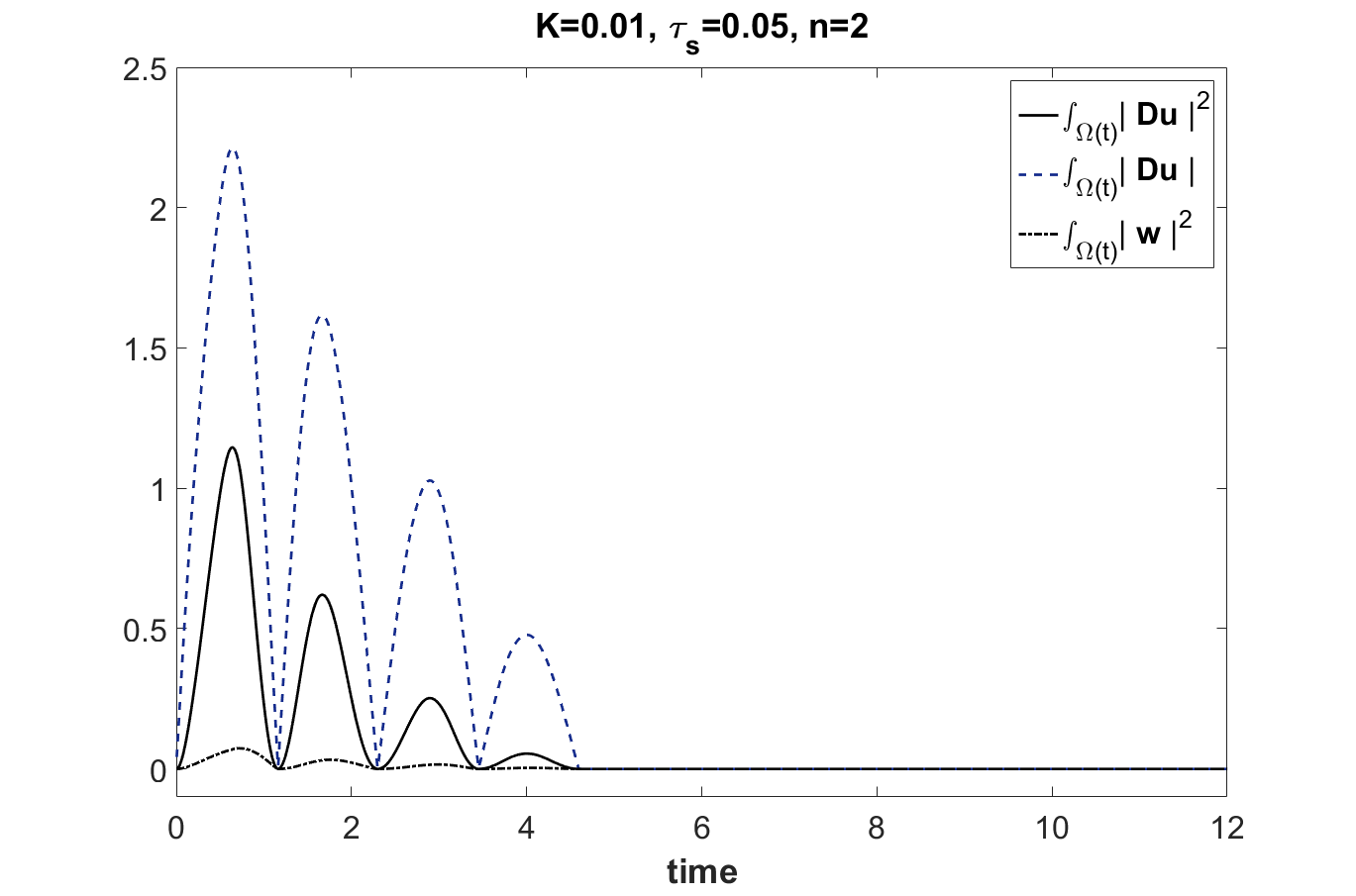}
 \caption{The evolution of $\int_{\Omega(t)}|\mathbf{D}\bu|^2\dx$ and of the enstrophy, $\int_{\Omega(t)}|\bw|^2\dx$, with $\bw=\frac{1}{\sqrt2}\nabla\times\bu$. All results are computed for  $n=2$ and different values of $\tau_s$.  For $\tau_s>0$ the figure also shows $\int_{\Omega(t)}|\mathbf{D}\bu|\dx$.
 \label{figDW}}
\end{figure}

Finally, we illustrate numerically the irrotational velocity field assumption (see discussion at the end of Section~\ref{s_energy}). 
To this end, we compare the skew-symmetric part  of  the velocity gradient tensor against its symmetric part for several computed solutions. For the skew-symmetric part we have $\int_{\Omega(t)}|\nabla_{scew}\bu|^2\dx=\frac12\int_{\Omega(t)}|\nabla\times\bu|^2\dx$.
Thus, Figure~\ref{figDW} shows the evolution of  $\int_{\Omega(t)}|\mathbf{D}\bu|^2\dx$ and of the enstrophy
for the  Newtonian droplet and for the yield stress case with two values of  parameter $\tau_s$.
For the yield stress fluid, we also plot $\int_{\Omega(t)}|\mathbf{D}\bu|\dx$, since this statistic enters the energy balance. In all cases, the produced vorticity appears to be minor compared to symmetric rate of strain tensor.
 }

\subsection*{Acknowledgements}
We are grateful to Kirill Terekhov (Stanford) for sharing the octree-CFD code and his help with setting-up numerical experiments. We also thank Roland Glowinski (University of Houston) for encouraging discussions.

\bibliography{refer}{}
\bibliographystyle{abbrv}

\end{document}